\documentclass[useAMS,usenatbib,a4paper]{mn2e}
\usepackage{rotating}       
\usepackage{graphicx}
\usepackage{amsmath} 
\usepackage{amssymb} 
\usepackage{natbib}

\title[Radio relic number counts and scalings]{How many radio relics await discovery?}

\author[Nuza et al.]{S.\,E.\,Nuza$^{1}$\thanks{E-mail: snuza@aip.de}, M.\,Hoeft$^{2}$,  R.\,J.\,van Weeren$^{3}$, S.\,Gottl\"ober$^{1}$ 
and G.\,Yepes$^{4}$\\
\\
$^1$ Leibniz-Institut f\"ur Astrophysik Potsdam (AIP), An der Sternwarte 16, 14482 Potsdam, Germany\\
$^2$ Th\"uringer Landessternwarte, Sternwarte 5, 07778 Tautenburg, Germany\\
$^3$ Leiden Observatory, Leiden University, P.O. Box 9513, NL-2300 RA Leiden, Netherlands\\
$^4$ Grupo de Astrof\'{\i}sica, Universidad Aut\'onoma de Madrid, Cantoblanco, 28039 Madrid, Spain\\
}

\date{}

\begin{document}

\maketitle

\begin{abstract}
  Upcoming radio telescopes will allow to study the radio sky at low frequencies
  with unprecedent sensitivity and resolution. New surveys are expected to
  discover a large number of new radio sources, in particular those with a steep
  radio spectrum. Here we investigate the abundance of radio relics, i.e.
  steep-spectrum diffuse radio emission coming from the periphery of galaxy
  clusters, which are believed to trace shock waves induced by cluster mergers.
  With the advent of comprehensive relic samples a framework is needed to
  analyze statistically the relic abundance. To this end, we introduce the
  probability to find a relic located in a galaxy cluster with given mass and
  redshift allowing us to relate the halo mass function of the Universe with the
  radio relic number counts. Up to date about 45 relics have been reported in
  the literature and we compile the resulting counts, $N(>S_{1.4})$. In
  principle, the parameters of the distribution could be determined using a
  sufficiently large relic sample. However, since the number of known
  relics is still small for that purpose we use the {\sc MareNostrum Universe}
  simulation to determine the relic radio power scaling with cluster mass and
  redshift. Our model is able to reproduce the recently found tentative 
  evidence for an increase in the fraction of clusters hosting 
  relics, both with X-ray luminosity and redshift, using an X-ray flux 
  limited cluster sample. Moreover, we find that a considerable fraction of faint 
  relics ($S_{1.4} \lesssim 10 \: \rm mJy$) reside in clusters with an X-ray flux below $ \lesssim 3
  \times 10^{-12}\: \rm erg \: s^{-1} \: cm^{-2}$. Finally, we estimate the
  number of radio relics which await discovery by future low frequency surveys
  proposed for the Low Frequency Array (LOFAR) and the Westerbork Synthesis
  Radio Telescope (WSRT). We estimate that the WODAN survey proposed for WSRT
  may discover 900 relics and that the LOFAR-Tier 1-120 MHz survey may discover
  about 2500 relics. However, the actual number of newly discovered relics will
  crucially depend on the existence of sufficiently complete galaxy cluster
  catalogues. 
\end{abstract}  


\section{Introduction}
\label{sec:intro}

  Some galaxy clusters show in their outskirts large-scale diffuse radio
  emission, which apparently does not originate from any individual galaxy. 
  These objects are called `radio relics'. Spectacular examples have been 
  found for instance in 
  A\,3667 \citep{roettgering:97}, A\,3376 \citep{bagchi:06}, and CIZA\,2242
  \citep{vanweeren:10}. Diffuse sources are difficult to detect due to the low
  surface brightness and due to the steepness of the spectra. Moreover, they can
  be only classified as relics if galactic foreground and fossil radio galaxy
  emission can be excluded and the hosting galaxy cluster is identified. We give
  a list of currently known relics in Section~\ref{sec:relic-list}.

  Radio relics show steep spectral slopes, which suggest that the origin of the
  radiation is synchrotron emission. Hence, radio relics indicate both the
  presence of relativistic electrons and magnetic fields. There are several
  approaches to estimate the strength of the magnetic field in the region of
  relics. Equipartition arguments have been applied leading to field strengths
  in the range $\sim0.5-2$ $\mu$G \citep{govoni:04}. For the northwest relic in A
  3667 upper limits of the hard X-ray flux in that region provide a lower bound
  for the magnetic field, namely $1.6$ $\mu$G \citep{nakazawa:09}. The Rotation
  Measure distribution of polarized emission from sources in the cluster volume
  or in the background allows to constrain the magnetic field strength and 
  spectrum leading to values of $\sim 1$--$5$ $\mu$G
  \citep{vogt:03,kuchar:09,2010A&A...513A..30B}.

  Magnetic fields in galaxy clusters are either primordial \citep{grasso:01} or
  injected in the protocluster region by AGN and/or galactic winds
  \citep{voelk:00}. Whatever the origin of the initial seed, some amplification
  mechanisms are required to account for their strength in clusters. Cosmological
  magnetohydrodynamics (MHD) simulations predict magnetic field strengths of the
  order of $\mu$G spread over the cluster volume \citep[e.g.][]{2005JCAP...01..009D}. These
  studies indicate that the amplification of the magnetic field resulting by
  pure adiabatic contraction is not sufficient to explain the observed magnetic
  field strengths. Merger events and accretion of material onto galaxy clusters
  are supposed to drive significant shear-flows and turbulence within the
  intra-cluster medium (ICM). This can in principle amplify 
  magnetic fields up to at least $\mu$G levels \citep[see][ and references
  therein]{dolag:08}.

  The morphology and temperature distribution of the X-ray emission of the
  clusters which host radio relics indicate that relics only occur in systems
  with an ongoing or recent merger, e.g. A\,754 \citep{2011ApJ...728...82M}. For A\,3667
  it has been shown that the relic is located where the bow shock of the moving
  sub-clump is expected \citep{vikhlinin:01}. For some other clusters the
  density and the temperature jump of the shock front at the position of the
  relic have been identified \citep[see e.g.][ and references
  therein]{markevitch:10}. This suggest the following scenario for the origin of
  the large-scale radio relics: cluster mergers lead to the formation of shock
  fronts which are responsible of electron acceleration causing the relic radio
  emission.

  Two main mechanisms for the acceleration of electrons have been proposed to
  explain radio relics: (i) adiabatic compression of fossil radio plasma by the
  shock wave \citep{ensslin:01,ensslin:02} or (ii) diffusive shock acceleration
  (DSA) by the Fermi-I process
  \citep{drury:83,blandford:87,jones:91,ensslin:98,malkov:01}. In the first
  scenario, radio relics should have toroidal and complex filamentary
  morphologies showing very steep, curved radio spectra due to inverse Compton
  (IC) and synchrotron losses. In the DSA scenario the electrons are accelerated
  by multiple crossings of the shock front (in a first order Fermi process)
  tracing shocks in the presence of ubiquitous magnetic fields. It is worth noting
  that other alternative scenarios have also been suggested
  \citep[e.g.][]{2010arXiv1011.0729K}.

  The formation of radio relics in cosmological simulations has been
  studied e.g. by \citet{2008MNRAS.391.1511H}, \citet{2009MNRAS.393.1073B} 
  and \citet{2010arXiv1006.3559S}. The latter studied structure formation shocks
  present in two cosmological boxes with a comoving volume of $64^3~h^{-3}$
  Mpc$^3$ and $200^3~h^{-3}$ Mpc$^3$, and use the non-thermal DSA radio model of
  \citet{2007MNRAS.375...77H} (hereafter HB07) to give an estimation of the
  radio relic luminosity function (RRLF) for $z=0$ and 1 at 1.4 GHz. On the
  other hand, \cite{2010A&A...509A..68C}, using a Monte Carlo approach, studied
  the ocurrence of `radio haloes' in merging galaxy clusters assuming that
  electrons are re-accelerated through MHD turbulence, posing interesting
  constraints for the upcoming Low Frequency Array (LOFAR) at 120 MHz. Here we
  would like to provide the appropiate scalings of the relic radiation as a function of 
  cluster mass and redshift for a given frequency, as well as to give some
  plausible predictions for upcoming radio surveys within the context of the DSA
  scenario.

  Currently, there are several new radio telescopes under construction, in
  particular, for the very low frequency regime. Moreover, several existing
  telescopes are getting significantly improved receivers or backends. For
  instance, in the Netherlands and neighboring countries LOFAR is almost
  completed. This instrument will survey the sky in the frequency range 
  from 30 to 240 MHz. New receivers and electronics of the expanded Very Large Array (eVLA)
  will drastically improve the sensitivity of the Very Large Array (VLA). 
  Furthermore, the Westerbork Synthesis Radio Telescope (WSRT) will 
  be equipped with focal-plane array receivers which will 
  be optimized for 1.4\,GHz observations. This will significantly increase the 
  field-of-view and will hence improve the survey speed tremendously rising the
  expected number of relic observations.

  In this work we develop a framework to relate the abundance of galaxy
  clusters in the Universe to the radio relic number counts. To this end we
  introduce the `radio relic probability density', i.e. the probability of
  finding a radio relic with a given radio power located in a galaxy cluster of given mass
  and redshift. A large sample of observed relics would allow to fully determine
  the probability density function for a given halo mass function in the
  Universe. However, since only about 45 relics are presently known we therefore use the 
  {\sc MareNostrum Universe} cosmological simulation as a way of determining how the probability
  density scales with cluster mass and redshift. In order to normalize this function 
  we compile a list of currently known radio relics. 

  From the resulting probability distribution we are also able to reproduce the recently 
  found fractions of clusters with relics in the combined NORAS$+$REFLEX cluster sample 
  presented by \cite{2011A&A...533A..35V}. 
  Finally, we draw some conclusions on the amount of non-identified relics due to the fact that the 
  hosting cluster is still not known. Moreover, we present some estimates on how many relics may be 
  identified by upcoming radio surveys assuming plausible survey specifications.
  
  This paper is organized as follows. In Section~\ref{sec:howto} we derive the
  formalism to estimate the number of radio relics and introduce the `radio
  relic probability function'. We also extend the usually assumed sharp transition for the 
  flux detection limit to a `discovery probability' which is more appropriate for relic samples. 
  In Section~\ref{sec:sim_model} we present the cosmological simulation used in
  this work and briefly summarize the shock detection method and the non-thermal
  radio emission model adopted. In Section~\ref{sec:obs_norm} we present the
  most up to date observed relic sample, discuss the normalization of our model
  counts and its comparison to observations and present our predictions for upcoming radio surveys. 
  Finally, in Section~\ref{sec:summary} we close the paper with the summary. 
  
  In what follows we assume a flat $\Lambda$CDM cosmology with a matter density parameter
  $\Omega_{\rm M}=0.27$, an amplitude of mass fluctuations $\sigma_8=0.8$ and a
  Hubble constant $H_0= 100~h$ km s$^{-1}$ Mpc$^{-1}$, with $h=0.7$
  \citep[e.g.][]{2011ApJS..192...18K}.

\section{How to estimate the number of observable relics?}
\label{sec:howto}

  In this Section we present the formalism aim at estimating relic number counts
  in a given radio survey as well as the radio relic probability density.

\subsection{Cumulative radio relic number counts}

  How many relics are seen in the sky above a given radio flux $S_{\nu}$ at the
  observing frequency $\nu_{\rm obs}$? The flux of a source with luminosity per
  unit frequency $P_{\nu} := \frac{{\rm d}P}{{\rm d}\nu}(\nu)$ located at
  redshift $z$ is given by    
\begin{equation}
  S_{\nu}(P_{\nu},z) =
  P_{\nu}
  \frac{1+z}{4\pi d^2_l(z)} 
  \label{eq:flux}
\end{equation}
  \noindent where $\nu$ is the rest frame frequency and $d_l(z)$ is the
  luminosity distance keeping in mind that the appropiate redshift correction
  for the frequency between the rest and observer frames needs to be considered.

  We introduce the luminosity function of  `radio relic clusters', i.e. the
  number of galaxy clusters per unit comoving volume and logarithmic relic radio
  power as a function of frequency and redshift 
\begin{equation}
  n_P(z) := \frac{{\rm d}^2 N}{{\rm d}V_{\rm c} \: {\rm d} \log P_{\nu} }(P_{\nu},z)
  \label{eq:nP}
\end{equation}
  \noindent where $N$ is the number of clusters and $V_{\rm c}$ the comoving 
  volume. The RRLF is obtained by the convolution of a halo mass function
  \citep[e.g.][]{2008ApJ...688..709T}
\begin{equation}
 n_M(z) := \frac{{\rm d}^2 N}{{\rm d}V_{\rm c} \: {\rm d} \log M}(M,z)
 \nonumber
\end{equation}
  \noindent with the `radio relic probability density' of finding a galaxy cluster of mass $M$, 
  redshift $z$ and relic radio power $P_{\nu}$. Therefore, the RRLF becomes
\begin{equation}
    \label{eq:RRLF}
    n_P(z)=\int_{-\infty}^{\infty} n_M(z) p(P_{\nu},M,z){\rm d} \log M	
\end{equation}
\noindent where the relic probability density $p(P_{\nu},M,z)$ fulfills the following condition
\begin{equation}
  \label{eq:prob_condition}
  \int_{-\infty}^{\infty} p(P_{\nu},M,z) {\rm d} \log P_{\nu} = 1.
\end{equation}
\noindent Hence, integrating Eq.~(\ref{eq:nP}) allows us to write the total abundance of relics per logarithmic flux bin as follows
\begin{equation}
 \label{eq:counts1}
 \frac{{\rm d} N}{{\rm d} \log S_{\nu}}(S_{\nu}) =
 \int_{0}^{\infty} n_P(z)
 \frac{{\rm d} V_{\rm c} }{{\rm d} z}(z) {\rm d} z
 .
\end{equation}
\noindent Note that we have used ${\rm d} \log P_{\nu}={\rm d} \log S_{\nu}$ since $P_{\nu}$ depends linearly on $S_{\nu}$.

In observations low luminosity radio relics are hard to identify since the surface 
brightness may be too low to exhibit typical morphological features or spectral index variations. 
Moreover, a diffuse radio object is only identified as a relic when the galaxy cluster can be 
unambigously detected. Depending on the mass of the cluster this may also be challenging. 
As a consequence, we introduce a `discovery probability', instead of a sharp flux-limit we use 
a smooth transition, which includes both the sensivity of the survey and the uncertanties present 
in the identification. We write this probability as  
\begin{equation}
\phi(S_\nu ) = \frac{1}{2}\left(1 + {\rm erf}\left(\frac{\log S_\nu -  \log S_{\nu}^{\rm eff}}{ w} \right)\right)
\end{equation} 
\noindent where the effective sensitivity, $S_{\nu}^{\rm eff}$, basically gives the flux-limit and $w$ the width of the transition.

  Finally, the cumulative radio relic function can be computed convolving
  Eq.~(\ref{eq:counts1}) with the `discovery probability' and multiplying 
  by the sky fraction, $f_{\rm s}$, covered by the radio survey 
\begin{equation}
N(>\log S_{\nu}) = f_{\rm s} \int_{\log S_{{\nu}}}^{\infty} 
\frac{{\rm d} N}{{\rm d} \log S_{\nu}}(S_\nu) \phi(S_\nu) {\rm d}\log S_\nu.
\label{eq:counts2}
\end{equation}
  \noindent The radio flux--luminosity relation given by Eq.~(\ref{eq:flux}) and
  the redshift integration of Eq.~(\ref{eq:counts1}) are fully determined by the
  cosmological parameters. Since recent cosmological observations show that the
  resulting parameters are well constrained the procedure described above can be
  considered as a direct relation between the radio relic probability density
  and the observed number counts.

\subsection{Radio relic probability density}
\label{sect:radio_extrapolation}

   If we were to build a perfect radio telescope that could detect even the
   faintest radio emission, which radio power distribution function linked to
   structure formation shocks should we expect? Merger shocks can persist in the cluster 
   periphery basically forever, hence, every cluster should show 
   some relic radio emission. On the other hand, very bright and very faint relics are most 
   likely rare events. We therefore expect that there is a typical 
   radio luminosity for a cluster with given mass and redshift although the related flux 
   is evidently below current detection limits. As a consequence, we 
   assume that the probability density to find a relic is given by a log-normal distribution:
\begin{equation}
    p(P_{\nu},M, z) 
    \propto
    \exp \left\{ - \frac{ (\log P_{\nu} - \log \bar{P_{\nu}})^2 } {2\sigma_P^2} \right\}
    \label{eq:prob}
\end{equation}
\noindent where $\sigma_P$ is the standard deviation of the logarithmic radio power and $\bar{P_{\nu}}$ is the {\it mean} radio power that scales with hosting cluster mass, observed frequency and redshift respectively. We parametrize this function as follows
\begin{eqnarray}
  \label{eq:logP_mean}	
  \log \bar{P_{\nu}} & = &
  \log P_0 + C_M \times\log\left(\frac{M}{10^{14.5}~h^{-1}~{\rm M}_{\sun}}\right)
  \\
  & + &
  C_z \times\log\left(1+z\right) + C_\nu \times\log\left(\frac{\nu_{\rm obs}}{{\rm 1.4 \, GHz}}\right){\rm .}
  \nonumber
\end{eqnarray}

  \noindent The radio power normalization is given by the `reference radio
  power' $P_0$ while the scaling with hosting cluster mass, redshift and
  observing frequency is governed by $C_M, C_z$ and $C_{\nu}$ respectively. Formally, a
  different functional form could have been chosen for the radio relic
  probability density as long as the condition given in
  Eq.~(\ref{eq:prob_condition}) is fulfilled. However, we will show in
  Section~\ref{sec:estimate_rrpf} that a log-normal function describes
  reasonably well the radio power distribution of our simulated relic samples.

\section{Simulating radio relics}
\label{sec:sim_model}

  In order to simulate radio relics we need to use a galaxy cluster sample
  extracted from a cosmological simulation and apply an emission and magnetic
  model to the present shock waves. In this section we present our cosmological
  simulation, the method used to detect shocked gas within the simulated volume
  and our radio power emission model. Finally, we estimate the parameters of the
  relic probability density using our cosmological simulation. 

\subsection{The simulated galaxy cluster sample} 
\label{sec:MN_clusters}

Our simulated galaxy cluster sample was selected from the {\sc MareNostrum Universe} 
cosmological simulation which is a non-radiative hydrodynamical run of a representative region of the 
Universe \cite[][]{2007ApJ...664..117G}.
The simulation was run with the smoothed particle hydrodynamics (SPH) {\sc Gadget-2} 
code \citep{2005MNRAS.364.1105S}. 
The adopted cosmology is in agreement with a flat $\Lambda$CDM scenario having a matter density 
parameter $\Omega_{\rm M}=0.3$, a baryon density parameter $\Omega_{\rm b}=0.045$, an intial matter power spectrum characterized by a scalar spectral index $n=1$ and normalized to $\sigma_{8}=0.9$, and a dimensionless Hubble parameter $h=0.7$. The simulation started at $z=40$ using a linear density field represented by $2\times1024^3$ gas and dark matter particles in a comoving box of 500 $h^{-1}$ Mpc on a side. 
The resulting mass resolution for gas and dark matter particles is 
$8.3\times10^9~h^{-1}$ M$_{\sun}$ and $1.5\times10^9~h^{-1}$ M$_{\sun}$ respectively.

Identification of bound structures is done using the parallel friends-of-friends (FoF) algorithm 
described in \citet{1999ApJ...516..530K} with a linking length of $0.17$ in units of the mean interparticle separation. 
In order to generate our galaxy cluster catalogs as a function of cosmic time we consider five different redshifts 
up to $z=1$, namely $z=0,0.25,0.5,0.75$ and $1$, and take the 500 most massive galaxy clusters present at each cosmic time. 
In this way, the range of cluster masses we are able to probe goes from $\sim10^{14}~h^{-1}$ M$_{\sun}$ 
up to $\sim2.5\times10^{15}~h^{-1}$ M$_{\sun}$, meaning that the baryonic component of the systems inside the virial radius 
is typically resolved with thousands of gas particles for the less massive clusters and with several tens of 
thousands for the most massive ones.

\subsection{Shock finding and radio emission in the simulation}
\label{sec:radio_model}

The cosmological SPH code {\sc Gadget} clearly accounts for shock
dissipation as shown by shock tube simulations
\citep[e.g.][]{2001NewA....6...79S}. To this end, artificial viscosity has
been introduced into SPH, which evaluates the local velocity field 
to estimate the dissipation \citep{1992ARA&A..30..543M}. However, this 
technique is not able to determine the Mach numbers, which are needed for combining
SPH simulations with parametric models for radio emission of relics.
Two methods have been introduced for locating shock fronts and
estimating their strength: \citet{2006MNRAS.367..113P} uses the
increase of entropy with time while \citet{2008MNRAS.391.1511H}
evaluates spatial entropy gradients in single snapshots of the
simulation. Here we apply a slightly modified version of the latter
approach.

Briefly, our scheme for locating shock fronts can be summarized as
follows. For a given gas particle we evaluate its pressure gradient
and define the {\it shock normal} of the particle as 
${\bf n}\equiv-\nabla{P}/|\nabla{P}|$. In case that the pressure gradient 
corresponds to a true shock front several conditions must be fulfilled. 
In particular, we demand that (i) the velocity field shows a negative divergence, (ii) the density increases from the upstream 
to the downstream region and (iii) that the latter is also valid for the entropy. Utilizing the Rankine-Hugoniot jump 
conditions for hydrodynamical shocks \citep[see e.g.][]{1959flme.book.....L} these requirements allow to determine 
the Mach number ${\cal M}_i$. For a conservative estimate we compute the Mach numbers according to all three conditions and then take 
the minimum. We wish to avoid the overestimation of the Mach number since this could lead to spurious strong radio emission. 
We apply this shock detection scheme to all gas particles inside a cube of size 10 $h^{-1}$ Mpc 
(comoving) centred in the centre of mass of the systems available in our FoF catalogs. We consider here merger 
shocks, i.e. shocks introduced by cluster mergers, which are found to have typical Mach numbers 
around $\sim2.5-3$ (Araya-Melo et al. submitted). We note that also fast galaxies in a rather cold ICM may generate shock fronts.

\begin{table}
\begin{center}
\begin{tabular}{lrrrrrr}
\hline
\hline
      & $\log P_0$  &  $C_M$  &  $C_z$  &  $C_{\nu}$ & $\sigma_P$ &\\
\hline
{\rm This work (`a')} & 21.35 & 2.56 & 3.43 & -1.20 & 0.85 \\
{\rm This work (`b')} & 21.53 & 2.22 & 2.49 & -1.15 & 0.85 \\
\hline
{\rm Skillman et al.} & 22.20 & 3.65 & 3.90  & - &  -   \\
\hline
\hline
\end{tabular}
\label{tab:parameters}
\caption{
         Best fit parameters for the radio relic probability density given by Eq.~(\ref{eq:prob}) using our set 
         of {\sc MareNostrum} clusters for magnetic field scaling models `a' and `b' (see text). 
         The reference radio power, $P_0$, is obtained in our models using available relic observations for 
         normalization (see Section~\ref{sec:normalization}). 
         As a comparison shown are the radio power scaling parameters obtained by \citet{2010arXiv1006.3559S} in the same 
         redshift range (they assume model `a' and an acceleration efficiency $\xi_{\rm e}=0.005$).  
        }

\end{center}
\end{table}

\begin{figure*}
\begin{center}
         \includegraphics[width=1.0\textwidth]{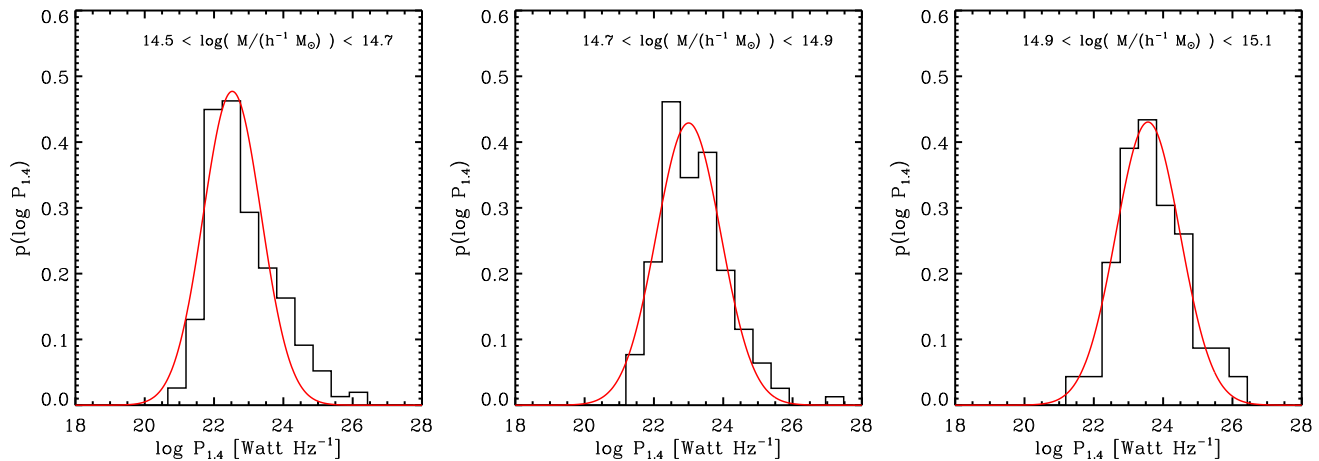}
	\caption{Radio power probability density for relics inside $3.6~h^{-1}$ Mpc from the cluster centre at $z=0$ and observing frequency $\nu_{\rm obs}=1.4$ GHz (magnetic model `a'; see text). The histograms show results from 
                 hosting clusters belonging to the {\sc MareNostrum} simulation for three different mass ranges and the solid lines 
                 are log-normal fits according to Eq.~(\ref{eq:prob}) normalized using the condition given 
                 by Eq.~(\ref{eq:prob_condition}).  
                }
	\label{fig:Lr-M-histo}
\end{center}
\end{figure*}

In order to predict the radio power of the simulated shock fronts we
need to know the magnetic field strength in the downstream area of the
shock fronts. Following our previous work in \citet{2008MNRAS.391.1511H} 
and that of \citet{2010arXiv1006.3559S} we assume
that the magnetic field is given by

\begin{equation}
  B = B_0 \times \left(\frac{  n_{\rm e}  }{ 10^{-4}\:{\rm cm^{-3}} }\right)^\eta
  \label{eq:B-n-scaling}
\end{equation}

\noindent where $n_{\rm e}$ is the local electron density, 
$B_0$ is a magnetic field reference value and $\eta$ is the slope of the density scaling. 
This dependence is motivated by the assumption that in average the magnetic field in the 
ICM is frozen in and that the gas motions distribute the magnetic field even to the outskirts 
of the cluster where luminous radio relics are generated. In fact, using Faraday rotation 
measures in the Coma cluster \citet{2010A&A...513A..30B} found evidence that magnetic 
fields are spread over the entire ICM. In this work we explore two magnetic models. 
In the first place we assume $B_0 = 0.1 \: {\rm \mu G}$ and $ \eta = 2/3$ 
\citep[e.g.][]{2008MNRAS.391.1511H} which typically leads to $\sim{\rm \mu G}$ values at the ouskirts 
of galaxy clusters (model `a'). We also adopt the 
scaling found by \citet{2010A&A...513A..30B} that produces higher magnetic field 
values ($\gtrsim {\rm \mu G}$) at these locations (model `b'). Their best-fit model 
indicates a slightly lower exponent than before but a stronger field for an electron density 
of $10^{-4}\:{\rm cm^{-3}}$, namely, $\eta = 1/2$ and $B_0=0.8\: {\rm \mu G}$, respectively.
It is worth mentioning that the upper limits for the IC emission in the hard X-ray band 
for the northwest relic in A 3667 indicate higher magnetic field strength 
\citep[$\ge 1.6 \: {\rm \mu G}$; ][]{nakazawa:09} than obtained here for a typical electron 
density of $\sim10^{-4}\:{\rm cm^{-3}}$. Hence, this could lead to an overestimation of 
the hard X-ray flux for a similar relic in the simulation. However, it is not currently known 
if the relic in A 3667 hosts an exceptionally strong magnetic field or if relics show 
in general field strengths of a few $\mu$G or more. 

As mentioned in the introduction the emission scenario (ii) states that thermal (or mildly relativistic) 
electrons are accelerated at the shock front by DSA. 
Important evidence for this mechanism comes from the relic in galaxy cluster CIZA~2242 \citep{vanweeren:10}.
In this case the observed gradient in the spectral index is consistent with electrons accelerated 
at the shock front, while synchrotron and IC losses cause the steeper spectral 
index in the downstream region.
The relic in CIZA~2242 also shows that radio emission originates from a rather
small volume in the downstream region of the shock with an extent less than 50~kpc. 
In HB07 we have worked out the relation between the radio emission and the 
properties of the shock front and downstream plasma within the context of the DSA model. 
Assuming that the relativistic electron population is advected with the 
downstream plasma and cooled down due to synchroton looses 
and IC scattering with CMB photons we are able to
estimate the total radio emission. In particular, the radio power per
unit frequency contributed by a SPH gas particle $i$ can be written as
follows
\begin{eqnarray}
     P_{\nu,i} 
     & = &
     6.4 \times 10^{34}\, {\rm erg \, s^{-1} \, Hz^{-1} } \;\;
     \frac{A_i}{{\rm Mpc^2}} \,
     \frac{ n_{{\rm e},i}}{\rm 10^{-4} \, cm^{-3}} \,
         \nonumber
     \\
     && \quad
         \times
     \frac{ \xi_{\rm e} }{0.05 } \:
     \left(\frac{\nu}{\rm 1.4 \, GHz} \right)^{-\frac{s_i}{2}}
     \left( \frac{{T_{{\rm d},i}}}{\rm 7\, keV} \right)^{\frac{3}{2}} \:
     \label{eq:radio_power}
     \\
     && \quad
     \times
     \frac{( B_{{\rm d},i} / {\rm  \mu G})^{1+\frac{s_i}{2}} }
              {(B_{\rm CMB} / {\rm \mu G} )^2 + ( B_{{\rm d},i} / {\rm \mu G} )^2 }
     \;
     \Psi({\cal M}_i)
     .
     \nonumber
\end{eqnarray}

\noindent In this formula, $A_i$ represents the surface area given by the SPH
particle, $n_{{\rm e},i}$ is the electron density, $\xi_{\rm e}$ is
the electron acceleration efficiency, $s_i$ is the shock compression
factor, $T_{{\rm d},i}$ is the post-shock temperature, $B_{{\rm d},i}$
is the post-shock magnetic field, $B_{\rm CMB}$ is the magnetic measure 
of the CMB energy density and $\Psi({\cal M}_i)$ is a function that depends
on the shock strength. 

We would like to note that the efficiency, $\xi_{\rm e}$, for electron
acceleration, denotes the fraction of the energy dissipated at the shock 
front that is transferred to supra-thermal particles. The lower energy 
threshold for supra-thermal particles is computed from the condition that 
the power-law distribution of supra-thermal electrons must meet the thermal 
electron distribution at the lower energy threshold, see HB07 for more details.
As a result of this approach the radio emission decreases drastically
for Mach numbers lower than 3. For the computations that follow we simply 
adopt $\xi_{\rm e}=0.005$. We encourage the reader to see HB07 for details.

\subsection{Estimation of the expected radio power scalings using the {\sc MareNostrum Universe}}
\label{sec:estimate_rrpf}

Our aim is to estimate the probability of finding relic radio emission coming from an arbitrary 
galaxy cluster having mass $M$, located at redshift $z$ and observed at frequency $\nu_{\rm obs}$. 
To this aim, we analyse the radio emission produced in the different clusters of our synthetic 
samples. As mentioned in Section \ref{sec:MN_clusters} we take the 500 most massive clusters at 
each considered redshift (i.e., $z=0,0.25,0.5,0.75,1$) and identify the shock fronts in each one of them. 
At all redshifts, we evaluate the radio power emitted from cluster relics as a function of 
hosting galaxy cluster mass and observing frequency. Note that when computing the radio power we consider 
all the luminosity caused by structure formation shocks within a distance of $3.6~h^{-1}$ Mpc (comoving) from 
the centre of mass without distinguishing between different relics. However, in each cluster, there are typically 
only one or two prominent relics which contribute most to the final radio emission.

We considered five different frequencies in our analysis, namely $\nu_{\rm obs}=0.12, 0.15, 0.21, 0.325$ and $1.4$ GHz. 
Fig.~\ref{fig:Lr-M-histo} shows the radio power distribution of relics at $z=0$ and $\nu_{\rm obs}=1.4$ GHz in the case 
of magnetic model `a', where the three panels show results for different hosting cluster mass. 
Best fit log-normal functions are also shown. We explore the parameter space of our relic cluster samples, 
given by the cluster mass, redshift and observing frequency, and repeat the fitting procedure 
of Fig.~\ref{fig:Lr-M-histo} (for simplicity we assume a constant value for $\sigma_P$). 
In this way, we are able to find a set of parameters for the radio relic probability 
density capable to reproduce the mean radio power scalings of our synthetic radio 
relics (see Eq.~(\ref{eq:logP_mean})). 
The best-fitting scaling parameters for the two magnetic field models adopted 
are shown in Table \ref{tab:parameters}. In general, the derived scalings show a good agreement. However, magnetic 
model `b' displays lower radio power scalings with mass and redshift in comparison with model `a', which is most noticeable 
for the redshift evolution. The reason for this can be understood in terms of the stronger magnetic field values achieved within the 
context of magnetic model `b'. Since according to Eq.~(\ref{eq:radio_power}) the radio emission saturates for large magnetic 
field values the resulting radio power scaling is not so pronounced in this case.

As can be seen from Fig.~\ref{fig:Lr-M-histo} log-normal functions reproduce reasonably well the radio power distribution of the 
synthetic relics. However, since there are possibly more small shocks, a better resolution in the simulation 
may serve to alleviate the observed skewness in the radio power distribution at low cluster masses. Additionally, 
this could let us extend the cluster mass range studied to estimate the mean radio power scalings. 
In the following section we present the currently known observed relic sample to further normalize our 
theoretical expectations with observations.

\section{How many relics do we expect?}
\label{sec:obs_norm}

\subsection{Compilation of currently known relics}
\label{sec:relic-list}

By definition, relics are diffuse radio emission in the periphery of
galaxy clusters without any optical counterpart. Hence, relics are
commonly searched for by correlating radio surveys with large
catalogues of galaxy clusters. The Westerbork Northern Sky Survey
(WENSS) has been carried out at 325\,MHz covering the north sky for declinations 
higher than 28.5$^\circ$. The noise level of this survey is 3.6\,mJy
\citep{rengelink:97}. The NRAO Very Large Array Sky Survey at 1.4\,GHz
(NVSS) covers the sky north of $-40^{\circ}$ and has a noise level of
0.45\,mJy \citep{condon:98}. Systematic searches for diffuse radio
emission in galaxy clusters have been undertaken, for instance, by
inspecting a sample of 205 X-ray bright Abell-type clusters in the
NVSS catalogue \citep{giovannini:99}, by analyzing the WENSS data at
the position of all Abell clusters \citep{kempner:01}, and by
searching for steep spectrum sources in the VLA Low-frequency Sky Survey 
\citep[VLSS;][]{2007AJ....134.1245C}
catalogue \citep{2009A&A...508...75V}.

\begin{figure}
\begin{center}
   \includegraphics[width=0.5\textwidth]{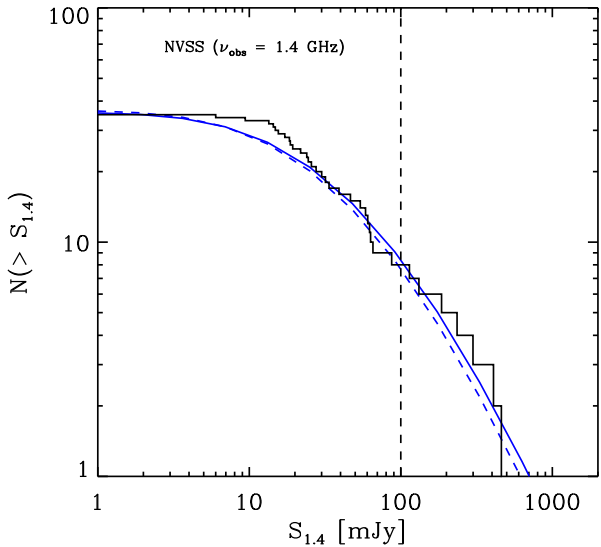}
   \caption{
            Cumulative number of NVSS radio relics. The histogram shows the 
            observed relic sample while the solid and dashed lines shows our magnetic 
            models `a' and `b' normalized to number counts at $S^{\rm eff}_{1.4}=100$ mJy 
            (vertical dashed line). The normalization leads to $\log P_0=21.35$ 
            and $21.53$ respectively (see Section~\ref{sec:normalization}). 
           }
   \label{fig:cum_obs_relics}
\end{center}
\end{figure}

As described above in more detail, current models for the formation of
relics are not able to \emph{predict} the actual number of observable relics by 
themselves, because both, the number density of relativistic electrons and the
strength of magnetic fields, are in general poorly constrained quantities. Therefore,
we wish to normalize the radio relic number counts, $N(>S_\nu)$, using 
the number of known radio relics. To this end, we have compiled a list of all
radio relics reported in the literature, as far as we are aware of, which 
can be seen in Table~\ref{tab:relics}. We have included all types of radio relics,
i.e. Mpc-scale single and double relics in the periphery of clusters
as well as smaller relics inside the cluster volume. A few of the small relics
might be attributed to the compression of fossil radio plasma \citep[known as 
`radio phoenix' class in the terminology of][]{2004rcfg.proc..335K}. 
However, we do not include phoenixes when computing the relic number counts.

\begin{figure*}
	\centering
        \includegraphics[width=1.0\textwidth]{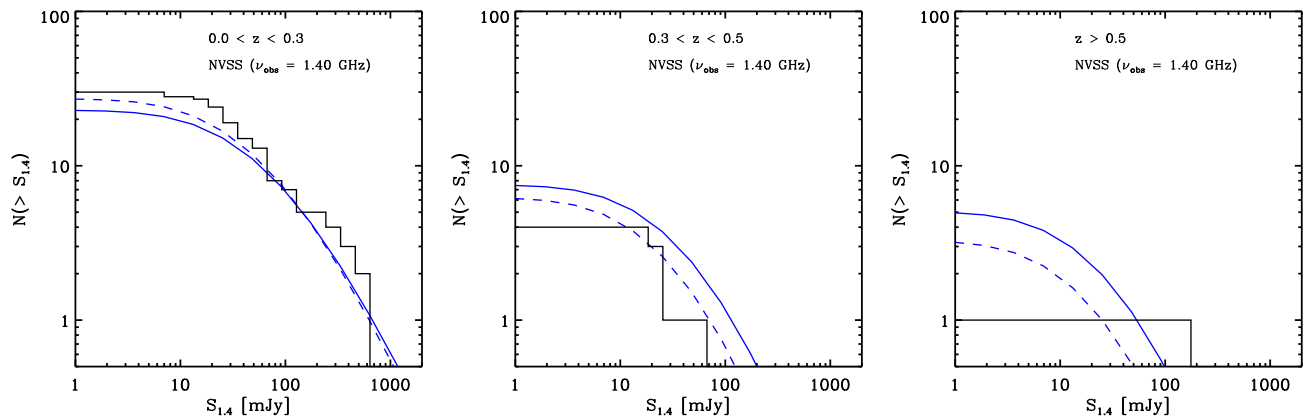}
	\caption{
                 Cumulative number of NVSS radio relics for different reshift bins. The histograms show results for the NVSS radio relic 
                 sample presented in this work, while solid (dashed) lines show the outcome of our magnetic model `a' (`b').
                 }
	\label{fig:model-z-bins}
\end{figure*}

For each cluster in Table~\ref{tab:relics} we give the flux of the
diffuse emission which has been classified as `relic' while the
contribution of radio haloes has been excluded. In the cases where halo
and relic radio emission are on top of each other due to projection effects
we only estimate the flux density of the relic emission. In many
clusters the diffuse relic emission is fragmented into multiple
pieces, e.g. A 2255 \citep{2008A&A...481L..91P}, or shows some prominent 
patches and very extended emission as well 
\citep[e.g. CIZA 2242;][]{vanweeren:10}. Instead of separating individual relics
in a single galaxy cluster, we combine the flux, $S_{\nu}$, of all relics in the
cluster which is consistent with defining the radio
luminosity probability for diffuse radio emission in clusters instead of that for relics 
(in the same way as done in Section~\ref{sec:radio_model}). For our analysis it is not useful 
to introduce relics as self-contained objects, since their identification depends 
inevitably on observational parameters such as sensitivity and resolution. Hence,
we give in column (3) of Table~\ref{tab:relics} the entire radio relic flux
present in each cluster. To normalize the relic number counts, $N(S_{\nu})$, we use 
the cluster flux at 1.4\,GHz because most of the measurements available are done 
in the 21 cm band. We also estimate the radio power of the relics 
at 1.4 GHz (see column (6) of Table~\ref{tab:relics}) assuming a spectral slope of -1.2 
which is consistent with the parameter $C_{\nu}$ given in Table~\ref{tab:parameters}.
Interestingly, A\,3667 displays an outstanding high flux. However, this object 
is not the most radio luminous relic as can be seen in Table~\ref{tab:relics}. 
The ten most luminous relics have fluxes $S_{1.4} \gtrsim 100\,{\rm mJy}$. It
is worth noting that several of these luminous relics have been
detected in recent years, namely 1RXS 06, CIZA 2242, and MACS
0717. On the other hand, the faintest relics known to date have a flux of $\sim 6$ mJy.

\begin{figure*}
	\centering
        \includegraphics[width=0.49\textwidth]{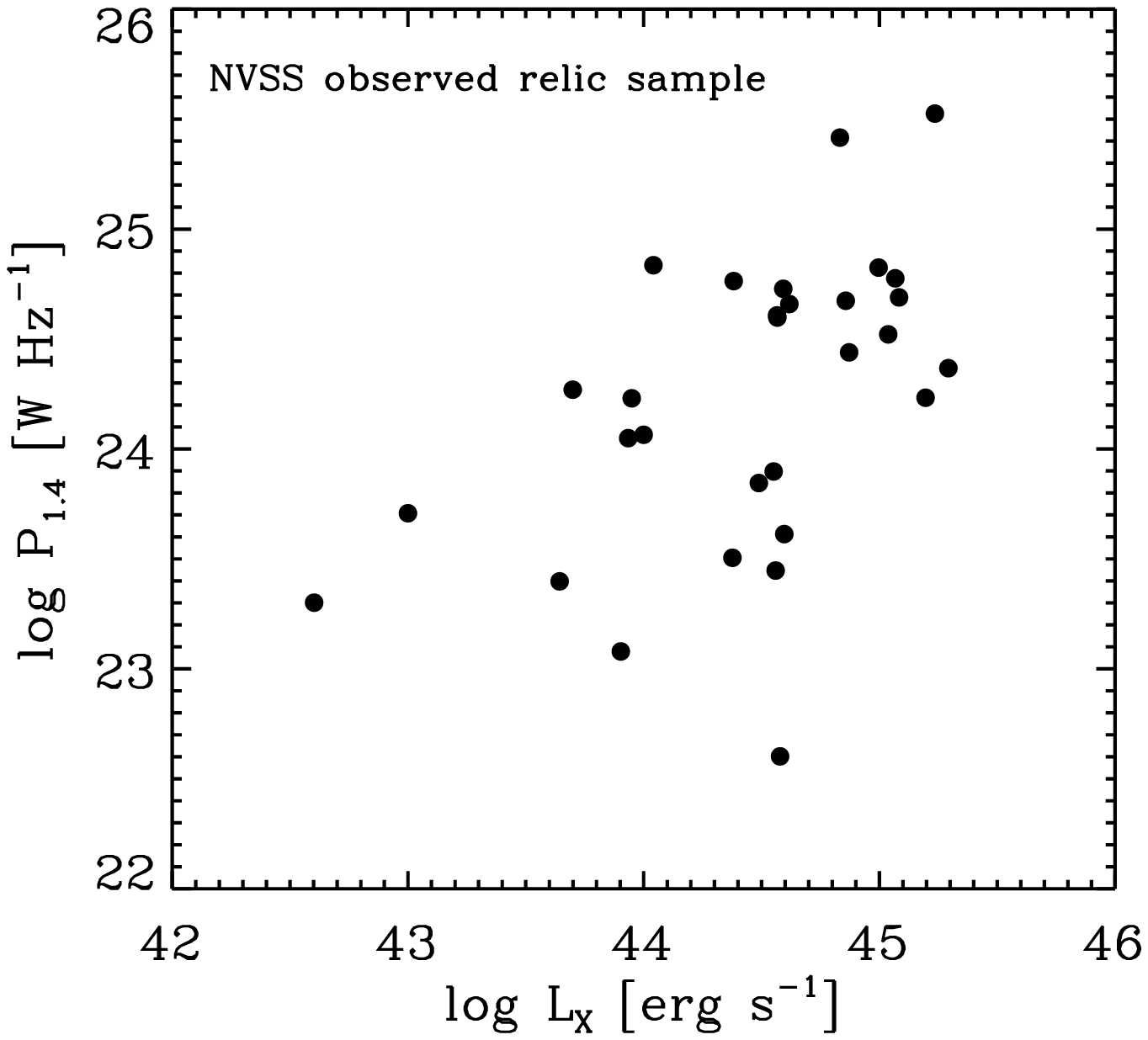}
        \includegraphics[width=0.49\textwidth]{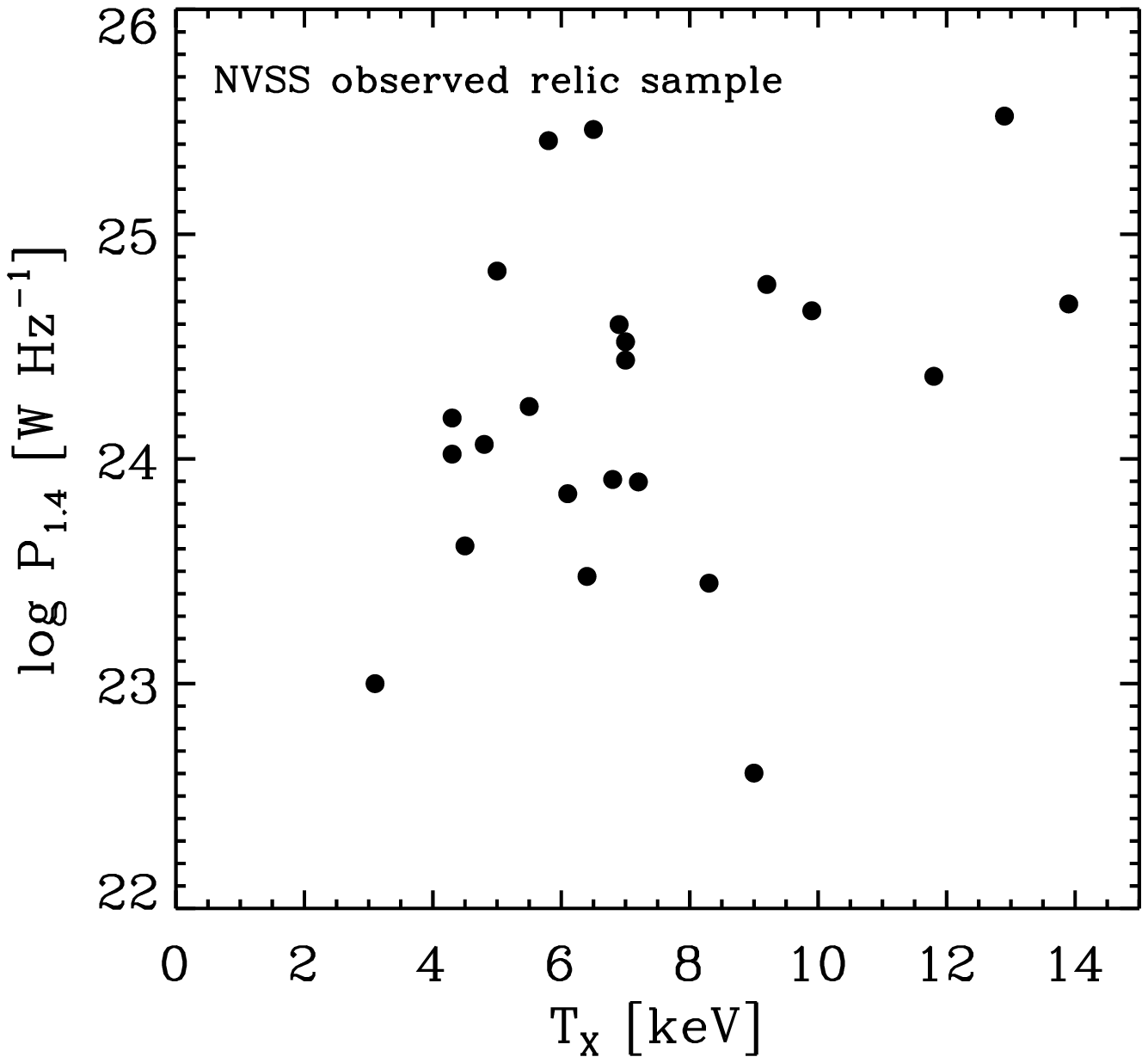}
	\caption{Estimated relic radio power at 1.4 GHz as a function of X-ray cluster luminosity (left panel) 
        and temperature (right panel) for the NVSS relic sample. All X-ray quantities are derived 
        in the ROSAT $0.1-2.4$~keV band (see Table~\ref{tab:relics}).  
        }
	\label{fig:P-Lx-Tx}
\end{figure*}

\subsection{Normalizing the radio relic number counts}
\label{sec:normalization}

Basically, we would like to normalize the predicted radio relic number
counts by using the bright-end of the observed number count distribution. 
As noted above amongst the ten brightest relics there are however three relics
which have been identified only recently. This could indicate that 
even the bright-end of the relics list does not contain all bright relic
sources on the sky which may introduce an offset in the global counts. 
Hence, in order to estimate number counts we assume a fiducial flux of 100\,mJy 
to centre the discovery probability. This means that at $S_{1.4}^{\rm eff} = 100\,{\rm mJy}$ 
half of the radio relic emission has been detected. Although this choice is arbitrary 
we take this value as a compromise between the lowest and brightest relics in the 
observed distribution. Furthermore, since the lowest flux of known radio relics is of a few mJy, 
we can set $w = 0.8$, which ensures that the discovery probability virtually vanishes below these 
values.

The non-detection of a relic could be due to several reasons. For 
instance, part of the sky might not be covered by deep radio surveys, 
galactic foreground radiation or bright sources in the cluster may 
obscure the diffuse emission, the surface brightness of the diffuse 
emission may be too low, or the related cluster could have not been identified 
yet. All these possibilities for the non-detection of existing diffuse radio 
patterns in a galaxy cluster are comprised in the complementary discovery 
probability $1-\phi$.

Having already a model for the relic discovery probability we are now able to
normalize the number counts to the present observed sample (which we dub as `NVSS'
since many candidates have been found by means of that survey). In order to do so 
we take from Table~\ref{tab:relics} all confirmed relics above a declination of $-40^{\circ}$ 
without including phoenixes. 
As can be seen in Fig.~\ref{fig:cum_obs_relics} the observed number of radio relics is well 
reproduced when using a normalization given by $\log P_0=21.35$ and $21.53$ for magnetic 
models `a' and `b' respectively. 
However, a degeneration between the normalization parameter and the detection 
threshold exists: a higher value for the normalization would imply a threshold higher than 100\,mJy 
if one is willing to reproduce observations. This would mean that the majority of 
relics with this higher flux has not been detected yet. We consider this possibility 
unlikely. On the other hand, we will show below that within the context of 
the magnetic models considered here a lower normalization can be ruled out as a result 
of the analysis of an X-ray flux limited cluster sample and their associated relics. The obtained low 
reference radio power ($\log P_0\sim21.4$) is enough to reasonably describe the observed 
distributions (see Section~\ref{sec:x-ray}).
Therefore, for the set of parameters derived above for the relic 
radio power probability ($C_M, C_z, C_{\nu}, \sigma_P$) we are able to constrain 
the normalization very well.

It is worth noting that to determine radio powers for the simulated clusters we 
had to {\it assume} an acceleration efficiency, $\xi_{\rm e} = 0.005$ 
(with $\xi_{\rm e}$ as defined in HB07 model), and we had to {\it assume} 
average scalings for the magnetic field, see Eq.~(\ref{eq:B-n-scaling}). 
For our {\sc MareNostrum} clusters these assumptions lead to a reference radio power of 
$\log P_0=22.23$ and $24.13$ in the case of magnetic models `a' and `b' respectively. 
Since, as mentiond above, to reproduce observations 
we require lower values for the normalization this implies that the acceleration efficiency must 
be $\xi_{\rm e} \lesssim 0.001$. In particular, for model `b' the observed acceleration efficiency 
could be about $\xi_{\rm e} \sim 10^{-5}$ which is more in line with theoretical expectations 
of DSA in Type Ia supernova remnants \citep[][]{2011MNRAS.414.3521E}. Further increase of 
the magnetic field values, as suggested by \citet{nakazawa:09}, would reduce the required 
efficiency even more.

Most of the observed radio relics have a redshift lower than 0.3. 
In the sample there are only five relics with higher redshift, and only one of them is 
located at $z>0.5$. We wish to compare these numbers to the predictions
according to the radio relic probability distribution. We simply split
the result into the redshift intervals given above, as can be seen in 
Fig.~\ref{fig:model-z-bins}. 
Apparently, our models predict more relics than observed for $z>0.3$. 
This might indicate that relics in distant clusters are more 
difficult to detect (e.g. due to resolution effects), that clusters would 
need to be more X-ray luminous to be found, or that 
our scaling parameter $C_z$ derived from the simulations does not agree 
with the actual redshift evolution. In particular, model `b' seems to better reproduce 
observations at all redshifts. However, it is important to realize that the number of 
both predicted and observed relics with $z>0.3$ is very small, so we should be
cautious with any interpretation. Much more extensive catalogues of
relics are needed to draw a significant conclusion about the redshift
evolution.

The most distant cluster which hosts a relic is MACS~0717.
Fig.~\ref{fig:model-z-bins} indicates that in our model the highest redshift relics 
should have fluxes $S_{1.4}$ within the range $10-50$ mJy. Instead, the relic in
MACS~0717 has a flux of about 140~mJy, indicating that this system is an 
outstanding radio relic. In fact, it is the most luminous relic known to date with 
$P_{1.4} \cong 2 \times 10^{26} \: \rm W \: Hz^{-1}$. 
For instance, using scalings resulting from magnetic model `a', the mean relic radio power of clusters having 
the mass and redshift of MACS~0717 is $\bar{P}_{1.4} \cong 4.1 \times 10^{24} \: \rm W \: Hz^{-1}$
(see Eq.~(\ref{eq:logP_mean})).
Hence, the luminosity of the relic is about $2 \sigma_P$ higher than the mean relic 
luminosity, so it is a rare event but still reasonably likely considering all clusters 
in the Universe.

In general, we can study the deviation between the estimated radio power from observations 
and the expected mean radio power at a given redshift and cluster mass. We can quantify this deviation 
in terms of $\Delta_P := \log\left(P_{1.4}/\bar{P}_{1.4}\right)/\sigma_P$, which measures 
the difference between the logarithmic radio power estimated from observations and the peak of the 
radio relic probability distribution in units of the parameter $\sigma_P$. In order to make a simple 
estimate we adopt the cluster X-ray luminosities given in column (8) of Table~\ref{tab:relics} 
to compute the cluster masses. In what follows we use a $L_X-M$ relation similar to that 
given by \citet{2009A&A...498..361P} that will be presented in Section~\ref{sec:fractions-x-ray-clusters}. 
In column (7) of Table~\ref{tab:relics} we give $\Delta_P$ for all relics in our observed sample adopting 
the scalings derived with the magnetic model `a' to estimate the mean radio power. 
As expected most of the relics display a mean deviation 
of $\sim 2\sigma_P$ since we are observing the brightest (or close by)
relics in the sky. Some of the relics show an unexpected high 
deviation from the mean radio power, $\bar{P}_{1.4}$, given by the hosting cluster 
mass and redshift. This may serve as an indication of the need to further investigate 
these systems in more detail to confirm their relic nature. However, the large deviations 
may also come from uncertanties in the cluster mass estimate. We do not pretend here to 
give a rigorous derivation of the hosting cluster masses but only assess the global tendency 
of the sample.

\subsection{The X-ray -- radio power relation}
\label{sec:x-ray}

\begin{figure}
	\centering
        \includegraphics[width=0.52\textwidth]{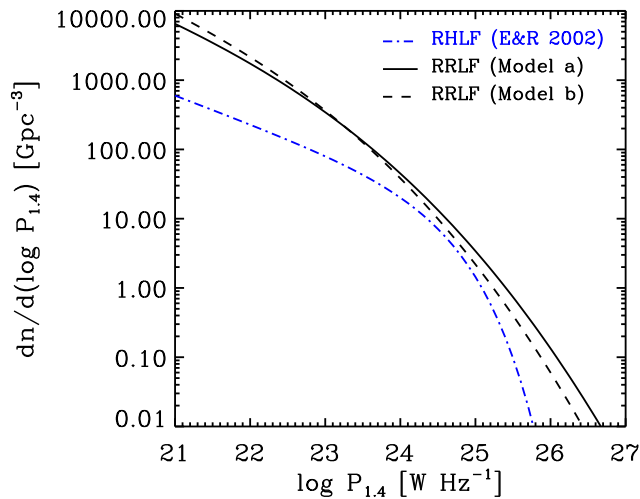}
	\caption{
                 Luminosity function of radio relics for models `a' and `b' 
                 (solid and dashed lines respectively) and radio haloes (dot-dashed line) at 1.4 GHz and $z=0$. 
                 The radio halo luminosity function is an analytic approximation 
                 taken from \citet{2002A&A...396...83E} under the assumption that 
                 a constant fraction $f_{\rm rh}=1/3$ of the clusters contain radio haloes. 
                }
	\label{fig:LF}
\end{figure}

For giant radio haloes in massive galaxy clusters a close correlation
between radio power and X-ray luminosity has been found
\citep[e.g.][]{2007A&A...463..937V}. In particular,
\citet{2002A&A...396...83E} suggested that the radio power of 
observed haloes scales with cluster X-ray luminosity according to
\begin{equation}
  \frac{ P_{\nu}^{\rm halo}(L_{\rm X}) }
       { {10^{24}\,{\rm W~Hz}^{-1}} }
  \:\: =  \:\:
   a_{\nu}
  \left(
    \frac{L_X}
    {10^{45}\,{\rm erg~s}^{-1}}
  \right)^{b_{\nu}}
  \label{eq:P-Xray-haloes}
\end{equation}
with parameters $a_\nu = 5.36$ and $b_\nu = 1.69$. In contrast, the
radio power of relics show a large scatter for a given X-ray
luminosity or temperature as can be seen in Fig.~\ref{fig:P-Lx-Tx}. 
This fact precisely reflects our starting point, namely, the 
recognition that the radio power of relics varies strongly for a 
given galaxy cluster mass. 
Therefore, this motivated us to introduce the radio power probability distribution.
Formally, we can relate the mean radio power, $\bar{P}_{1.4}$, to the
X-ray luminosity of clusters at $z=0$ using the $L_X-M$ relation 
in Eq.~(\ref{eq:logP_mean}) for the parameters 
given in Table~\ref{tab:parameters}. Comparing this result with 
Eq.~(\ref{eq:P-Xray-haloes}) we get for 
$a_\nu$ and $b_\nu$ the values $1.3\times10^{-3}$ and $1.5$, respectively. 
Interestingly, we find a similar exponent but a much lower 
proportionality constant. This seems to indicate that there are much 
less bright radio relics than haloes. 
However, we have to keep in mind that when computing the RRLF not only 
the {\it mean} radio power but the radio power distribution function 
needs to be taken into account, see Eq.~(\ref{eq:RRLF}). 
As a consequence we expect more radio relics than haloes as 
can be seen in Fig.~\ref{fig:LF}.

\subsection{Radio relics and an X-ray selected cluster sample}
\label{sec:fractions-x-ray-clusters}

The radio power probability density introduced above allows us to predict
the fraction of clusters in an X-ray selected sample which host a radio relic. 
As a first step, we introduce the differential distribution with respect to cluster X-ray flux
\begin{equation}
  n_{S_X S_\nu}
  :=
  \frac{ {\rm d}^2 N }
       { {\rm d}  \log S_X \: {\rm d} \log S_\nu }
  .
\end{equation}

\begin{figure}
        \vspace{0.5cm}
	\centering
        \includegraphics[width=0.5\textwidth]{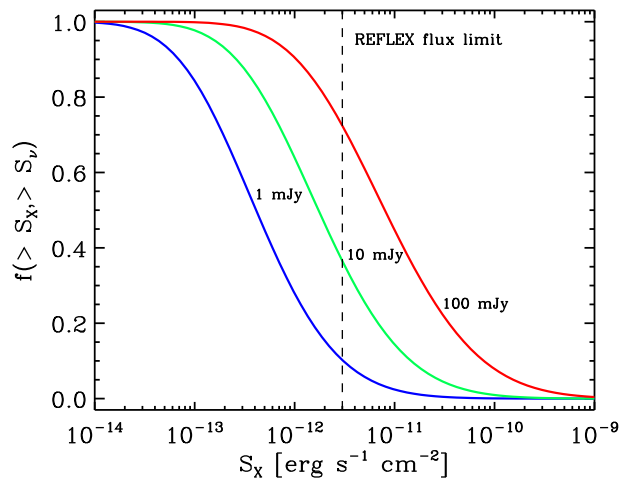}
	\caption{ 
        Cumulative fraction of clusters which host diffuse relic emission. The
        cumulative number is depicted as a function of X-ray flux, measured in
        the ROSAT $0.1-2.4$~keV band. The curves give the cumulative number
        for a relic flux of 1, 10, and 100~mJy. In addition the completeness
        limit of the REFLEX cluster sample is given. 
        }
	\label{fig:cum_Sx}
  \end{figure}

\noindent In a similar way as done in Eq.~(\ref{eq:counts1}) we can write

\begin{equation}
  n_{S_X S_\nu}
  =
  \int_0^\infty
  n_M (z) \:
  p( P_\nu, M, z ) \:
  \frac{{\rm d}  \log M}
       {{\rm d}  \log S_X }  \:
   \frac{{\rm d} V_{\rm c}}
       {{\rm d} z}  \: {\rm d}z
  .
  \label{eq:dNdSxdSnu}
\end{equation}
Note that radio power is a function of radio flux and redshift,
similarly, the X-ray luminosity is a function of X-ray flux and
redshift. Using the $L_X-M$ relation (see below) to estimate the 
cluster mass, based on its X-ray luminosity, we can write
$p( P_\nu, M, z ) = p( P_\nu(S_\nu,z), M(S_X,z), z )$. We now
introduce an X-ray flux threshold, $S_X^{\rm th}$ and assume that only
clusters with a flux above the threshold are detected. This allows us
to determine the cumulative fraction of clusters with $S_X > S_X^{\rm
th}$ that host diffuse relic emission with a given flux $S_\nu$.
Integrating the previous equation leads to
\begin{equation}
  F_X(>S_X^{\rm th},S_\nu)
  =
  \frac{1}{N_X(S_\nu)}
  \int_{S^{\rm th}_{X}}^\infty n_{S_X S_\nu} \:
  {\rm d} \log S_X
  ,
  \label{eq:f_X}
\end{equation}
where the normalization factor, $N_X(S_\nu)$, is determined by the
condition $F_X(>0,S_\nu)=1$. Fig.~\ref{fig:cum_Sx} shows the 
cumulative fraction of clusters for three different radio fluxes 
(for the sake of simplicity we assume only the radio power scaling 
parameters that result from model `a' throughout this section; adopting those 
of model `b' do not modify our main conclusions). 
About 80\% of the clusters which host diffuse relics with 100~mJy have
an X-ray flux larger than $3\times 10^{-12} \, \rm erg \, s^{-1} \,
Hz^{-1}$, which corresponds to the completeness limit of the REFLEX cluster 
sample \citep{2004A&A...425..367B}.
Since candidate radio relics are commonly identified by 
cross-correlating radio and X-ray catalogues, the NORAS
and REFLEX cluster catalogues are well suited to identify luminous
relics. In contrast, for faint relics of about 10~mJy only $\sim 40
\%$ of the hosting clusters are expected to have fluxes above the 
REFLEX X-ray flux limit. This means that, if upcoming surveys
will allow the detection of diffuse radio structures with fluxes of
about 1~mJy, significantly deeper X-ray cluster catalogues will be
needed to identify the majority of radio relics.

\begin{figure}
	\centering
        \includegraphics[width=0.5\textwidth]{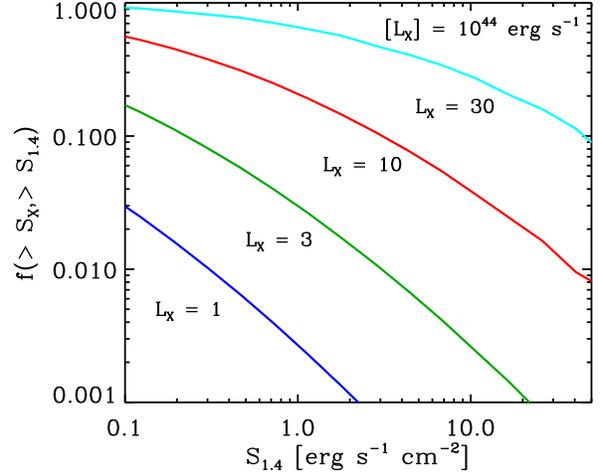}
	\caption{ 
        Cumulative fraction of clusters with relics as a function of radio flux. 
        The cumulative number is depicted for four different X-ray luminosities.  
        }
	\label{fig:cum_Snu}
  \end{figure}

In a similar way we can determine the cumulative fraction of clusters
which host diffuse relic emission as a function of radio flux. To this
end we introduce the fraction of galaxy clusters with detectable relics at a
given mass and redshift
\begin{equation}
  f_\phi(M,z)
  =
  \int_{-\infty}^{\infty} p( P_\nu, M, z ) \:
  \phi(S_\nu) \: {\rm d} \log P_\nu
  ,
\end{equation}
\noindent where $\phi$ is the discovery probability introduced earlier. 
Note that for $w \rightarrow 0$ we can mimic a 
Heaviside-function, i.e. only relics above $S_\nu^{\rm eff}$ are
detected. We can now determine the cumulative fraction of clusters
hosting radio relics per X-ray luminosity bin, $\Delta \log L_X$, as follows
\begin{equation}
    F_\nu(>S_\nu^{\rm eff}, L_X) =
    \frac{1}{N_\nu(L_X)}
    \int_0^\infty n_{M,f}
    \frac{{\rm d}  \log M}
    {{\rm d}  \log L_X }
    \frac{{\rm d} V_{\rm c}}
    {{\rm d} z} \:
    {\rm d}z   
    \label{eq:f_nu}   
\end{equation}
\noindent where $n_{M,f} := n_M\:f_\phi(M,z)$ and the normalization factor,
$N_\nu(L_X)$, is given by the condition $F_\nu(>0, L_X) = 1$.
Fig.~\ref{fig:cum_Snu} shows the cumulative fraction for different
cluster X-ray luminosities. As expected, only a small fraction of
clusters show radio relics for current detection limits of about
10~mJy. The fraction decreases strongly with the cluster X-ray
luminosity. For instance, $\sim20 \%$ of clusters with an X-ray luminosity 
of about $3\times10^{45} \: \rm erg \, s^{-1}$ are expected to host a 
relic with a flux of 10~mJy or brighter, while only $\sim0.3 \%$ of clusters 
with $3\times10^{44} \: \rm erg \, s^{-1}$ are expected to do so.
Note that in Eq.~(\ref{eq:f_nu}) we have assumed that {\it all}
clusters with given X-ray luminosity can be detected. To compare the
result to X-ray selected cluster samples we need to introduce an X-ray flux
limit. As a result, a large number of faint relics residing in
distant clusters falls below the flux limit.

In a recent work \citet{2011A&A...533A..35V} selected 544 clusters from the
NORAS \citep{2000ApJS..129..435B} and the REFLEX
\citep{2004A&A...425..367B} cluster samples with an X-ray flux above
$3\times 10^{-12} \: \rm erg \, s^{-1} \, cm^{-2}$ and located 
outside the galactic plane. Up to this flux the 
REFLEX sample is virtually complete. 
On the other hand the NORAS sample is almost 50\% complete. 
Interestingly, these authors show that 16 out of 
the 544 clusters of the combined list contain at least one radio relic
and found evidence for an increase of the fraction of clusters which
host relics with cluster X-ray luminosity and redshift.

Eq.~(\ref{eq:f_nu}) allows us to determine the fraction of clusters with 
relics. The X-ray flux limit imposes an upper limit in the redshift integral, 
i.e. allowed redshifts must fulfill the condition $z<z(S_X^{\rm th},L_X)$. 
First we wish to reproduce the ${\rm d} N_{\rm cl} / {\rm d } L_X$ 
and ${\rm d } N_{\rm cl} / {\rm d} z$ distributions of the cluster 
sample selected in \citet{2011A&A...533A..35V}. To this end we rewrite 
Eq.~(\ref{eq:f_nu}) as follows
  \begin{equation}
    \frac{{\rm d} N_{\rm cl}}{{\rm d } L_X}
    =
    f_{\rm s }
    \int_0^{z_{\rm th}}  \,
    n_M \:
    \frac{{\rm d}  \log M}
         {{\rm d}  L_X } \,
    \frac{{\rm d} V_{\rm c}}
         {{\rm d} z} \:
    {\rm d}z
    \label{eq:dNdLx}
  \end{equation}
  and
  \begin{equation}
    \frac{{\rm d} N_{\rm cl}}{{\rm d } z}
    =
    f_{\rm s }
    \int_{L^{\rm th}_{X}}^\infty \,
    n_M \:
    \frac{{\rm d}  \log M}
         {{\rm d}  L_X }
    \frac{{\rm d} V_{\rm c}}
         {{\rm d} z} \:
    {\rm d} L_x
    ,
    \label{eq:dNdz}
  \end{equation}

\begin{figure*}
   \vspace{1cm}
   \includegraphics[width=0.4\textwidth]{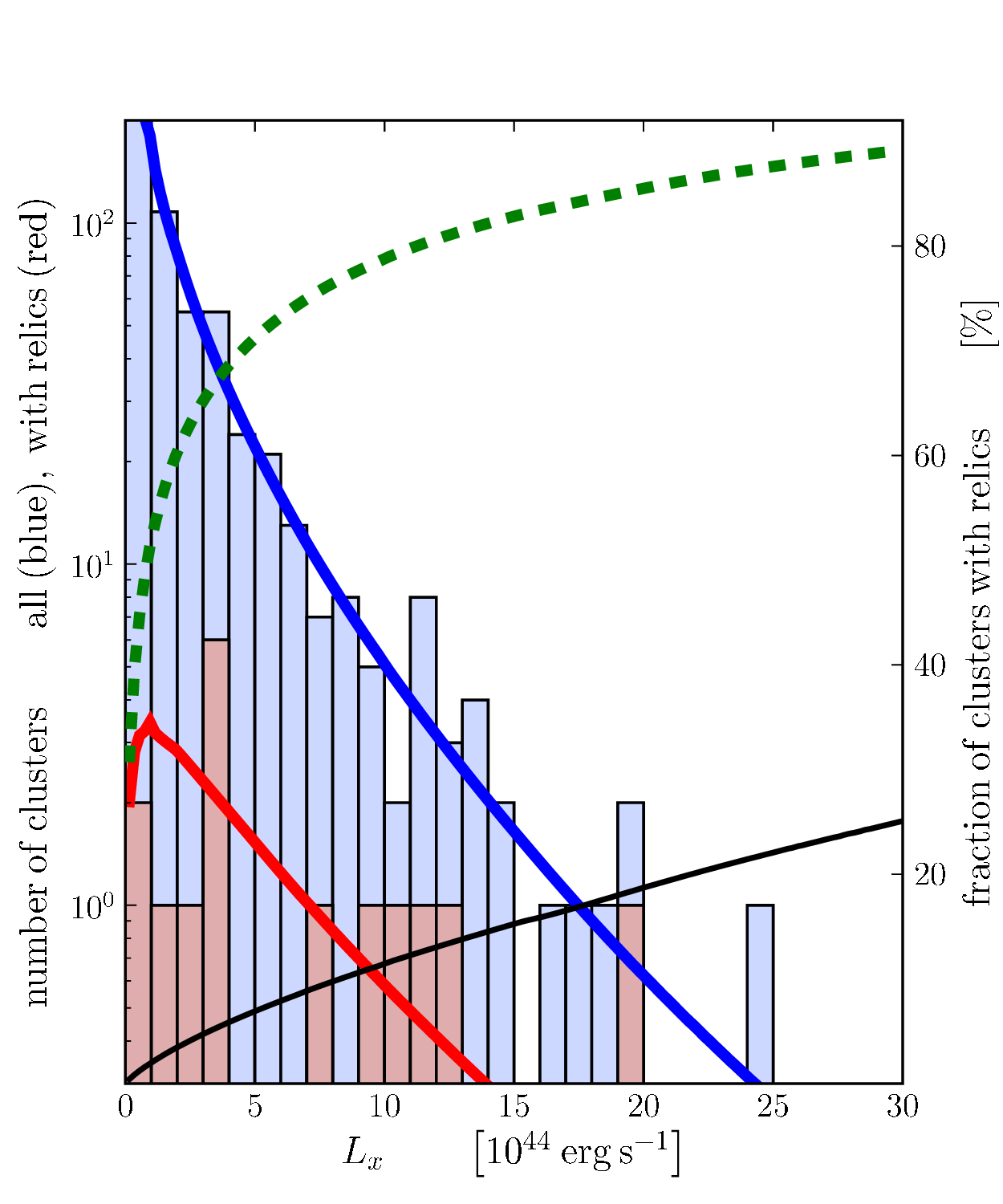}
   \hspace{1cm}
   \includegraphics[width=0.4\textwidth]{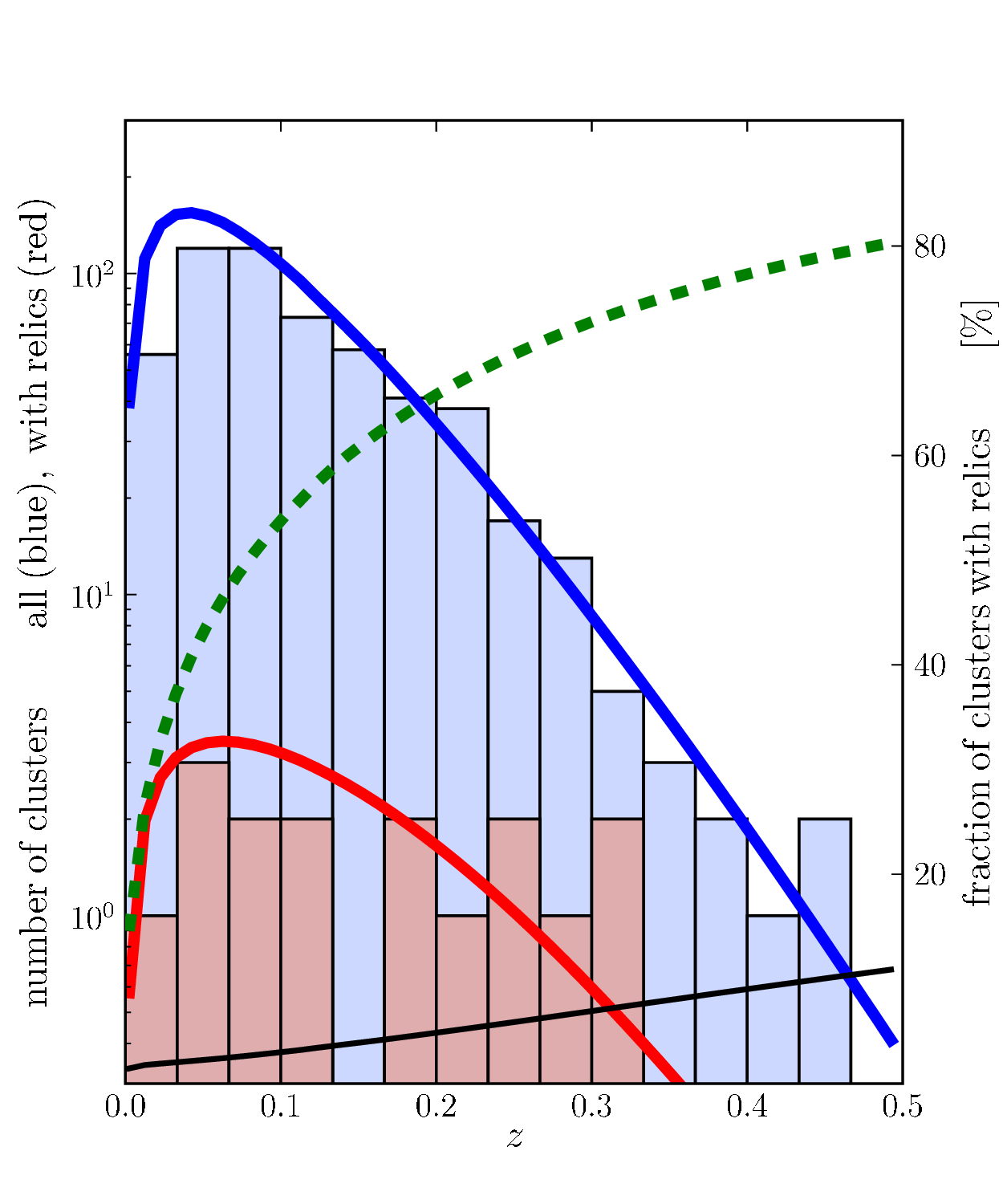}
   \caption{
     Luminosity (left panel) and redshift (right panel) distributions of X-ray
     clusters in the NORAS$+$REFLEX sample with and without radio relics (red and
     blue histrograms respectively). Blue and red solid lines show the
     distributions expected in our model while the black solid line shows the
     fraction of clusters with relics (ratio between the red and blue solid
     lines). The green dashed lines show the expected fraction of clusters with relics 
     for the NORAS sample in the upcoming LOFAR-Tier~1-120 MHz (see Section~\ref{sec:predictions}).
           }	
   \label{fig:cluster-fractions}
\end{figure*}

 \noindent where $N_{\rm cl}$ is the number of clusters and $f_{\rm s}$
  indicates the sky fraction covered by the selected cluster sample,
  we estimate it to be 35\%\footnote{The sky fraction is estimated by 
  $f_{\rm sky} \times f_{\rm int} \times (1-f_{\rm gal})$, where $f_{\rm sky}$ is 
  the sky fraction covered by the survey which overlaps NVSS, $f_{\rm int}$ is the 
  completeness of the survey, and $f_{\rm gal}$ is the fraction of clusters located 
  at a galactic latitude lower than $20^\circ$. For NORAS and REFLEX we assume 
  50\%, 50\% and 33\%, and 34\%, 90\%, and 33\%, for these quantities respectively. 
  The sum of the two contributions leads to the estimated value.}. The integration boundaries, $z_{\rm
  th}(S^{\rm th}_{X},L_X)$ and $L^{\rm th}_{X}(S^{\rm th}_{X},z)$ are
  obtained from the flux limit in the survey, see Eq.~(\ref{eq:flux}). 
  To perform the integration we need to relate cluster mass and X-ray luminosity. 
  We assess the cluster luminosity and redshift 
  distributions by choosing an appropriate $L_X(M)$ relation.
  Recently, \citet{2009A&A...498..361P} investigated cluster scaling
  relations in detail. They provide best fit parameters for $L_{X,500} -
  M_{500}$, where both quantities are measured within $R_{500}$. However, our
  sample differs in several respects to theirs: 
  (i) the halo mass function, $n_M$, is given for $M_{200}$, (ii) we do not extrapolate the
  luminosities to $R_{500}$, (iii) our sample extends up to $z \sim
  0.5$ while \citet{2009A&A...498..361P} use only clusters with
  $z<0.2$, and (iv) we use a slightly different value for $\Omega_{\rm M}$.
  Adopting the scaling relation
  \begin{equation}
    \frac{L_X}
         {5 \times 10^{43} \, \rm erg \, s^{-1}}
    =
    E(z)^{7/3}
    \:
    \left(
       \frac{M_{200}}
            {3.4 \times 10^{14} \, h^{-1} \, {\rm M}_\odot }
    \right)^{1.55}
    \label{eq:Lx-M}
  \end{equation}

\noindent with $E(z)=\sqrt{\Omega_{\rm M}(1 + z)^3 + \Omega_{\Lambda}}$ we are able to reproduce 
the observed distributions reasonably well (see Fig.~\ref{fig:cluster-fractions}).

To compute the abundance of clusters which actually hosts a detectable
relic we have to use $n_{M,f} = n_M f(M,z)$, instead of $n_M$ in
Eqs.~(\ref{eq:dNdLx}) and (\ref{eq:dNdz}). Fig.~\ref{fig:cluster-fractions}
(red solid lines) shows the resulting X-ray luminosity and redshift
distributions. 
Note we use here a lower effective sensitivity and smaller 
width in $\phi(S_\nu)$ because the clusters are {\it known}. As discussed
above we may miss the discovery of a bright relic, since the cluster has not 
yet been identified. For instance, the double relic in PLCK G287.0 has been 
discovered only after the detection of the hosting cluster with {\sc Planck} satellite even if the 
diffuse emission is clearly visible in NVSS. Since we consider here the relics in the
NORAS$+$REFLEX sample all clusters are known by construction. We model the discovery probability in the following way: (i) we argue that the width is smaller than
for the overall sample, we take $w=0.2$, (ii) we adjust $S^{\rm
eff}_{1.4}$ to reproduce the fraction of clusters with relics found in the
sample, using $S^{\rm eff}_{1.4} = 27 \: \rm mJy$ we obtain a fraction of
3\%. 
Hence, the effective sensitivity adopted corresponds to $60$ times the r.m.s. 
noise level in the NVSS survey. In this way we find that the fraction of
relics in an X-ray flux limited cluster sample should indeed
increase with X-ray luminosity and redshift as shown in Fig.~\ref{fig:cluster-fractions}. 
The redshift distribution is a crucial test for the mean radio power scaling 
parameter $C_z$.

\subsection{Predictions for upcoming surveys}
\label{sec:predictions}

In Table~\ref{tab:surveys} we have summarized specifications for planned surveys with LOFAR and the WSRT. We would like to give some 
plausible estimates for the expected number of relics to be discovered by upcoming radio surveys. However, we have to remember that there 
may be several uncertainties difficult to quantify. 
As we already mentioned the determination of $p(P_{\nu},M,z)$ is affected 
by the limitations of our simulation and by the adopted physical model used to relate the Mach number with 
the relic radio power. However, since we are able to reproduce the trends found for the NORAS$+$REFLEX
sample we conclude that our approach has resulted in a reasonable set of parameters.
In the present paper we use the radio relic probability density 
estimated from the {\sc MareNostrum} simulation and leave for future work a more extensive modelling 
of the radio power emission.

To compute number counts for future surveys we also need to assess the discovery probability 
for each survey.   
Since it is beyond the scope of this work to model this in detail we simply adopt 
a conservative approach: we assume that the NVSS detection parameters hold 
similarly for the upcoming surveys, i.e. we take $w=0.8$ (detection/non-detection transition of the instrument) 
and $b := S_{\nu}^{\rm eff}/\sigma_{\nu} \sim 200$ 
(ratio of the effective sensitivity of the overall relic sample to the survey noise) as our fiducial parameters. 
The last condition means that for half of the clusters which host diffuse radio emission with a flux above 
200 times the noise level of the survey a radio relic will be detected. 
At this point, an important remark needs to be done in relation to the sensitivity per beam achieved 
by a given radiotelescope. The next generation of radio surveys will presumably increase their beam resolution 
at least by a factor of a few. In principle, for some of the relics, this could imply the requirement of higher 
$b$ values than assumed here which may lead to an overestimation of the predicted number counts. For instance, in its final configuration the 
LOFAR telescope is expected to reach an angular resolution of $\sim 5$ arcsec. However, since this instrument has many 
short baselines it is always possible to smooth images down to typical NVSS resolution values (i.e. $\sim45$ arcsec) 
without increasing too much the resulting r.m.s. sensitivity. 
Nevertheless, we have to keep in mind that the final predicted number counts will depend on the adopted 
radio power scalings and detection parameters. 
For instance, in the case of the LOFAR-Tier 1-120 MHz survey, if we let the effective 
detection threshold to vary in the range $b=150-300$, keeping the remaining parameters fixed, 
our predictions will increase (decrease) by a factor of two (a half) with respect 
to the prediction corresponding to $b\sim200$. Similarly, one could assess the impact of 
varying some of the radio power scaling parameters keeping the rest unchanged. In particular, 
if we let the slope of the scaling with cluster mass to vary in the range $C_M = 1.5-3.5$ we 
get brighter (fainter) relics located in clusters with masses 
below (above) $10^{14.5}~h^{-1}~{\rm M}_{\sun}$. The predictions in this case will 
also increase (decrease) in a similar amount as before. 

For calculation purposes here we use the radio power scalings derived from the simulation 
assuming the fiducial detection paramaters presented above.
Table~\ref{tab:surveys} gives the total number of expected relics up to $z=0.3$, 
as well as relics with $0.3<z<0.5$, $0.5<z<1$, and $z>1$ for magnetic models `a' and `b' 
(in brackets). For $z>1$ models with higher magnetic fields would generally 
produce less radio relics as a consequence of the resulting redshift evolution.
Under these assumptions we expect that the LOFAR-Tier~1-120 MHz survey and WODAN 
large sky coverage survey should reveal several thousands of radio relics, due to the huge 
improvement in sensitivity that will presumably be achieved by these surveys. 
However, a given survey may provide candidates for radio relics, by cross-correlation 
with positions of known galaxy clusters. This means that deep follow up observations may 
be needed to confirm radio relics in the clusters. Interestingly, within the context of 
our simple scaling for the magnetic field, we also expect a significant number of relics 
with $z>0.5$ and $z>1$ to be detected. The actual number of relics at high redshift will 
in particular serve to constrain the redshift evolution of magnetic fields in clusters 
allowing us to further refine our model prescriptions. 

As noted above, the number of relics detected in upcoming surveys 
will crucially depend on the cluster database used to correlate candidates 
displaying diffuse radio emission with known cluster positions. As a consequence, 
the final number of unambigously identified relics could be below our model 
expectations. To quantify this we estimate how many relics LOFAR should find 
in the Tier~1-120~MHz configuration for the NORAS cluster sample introduced before. 
Note that the LOFAR survey and the NORAS sample cover both the north sky.
Based on the effective sensitivity found in the previous 
section for relics in this galaxy cluster sample we adopt here 
a flux threshold $60$ times that of the r.m.s. noise level of the 
LOFAR-Tier 1-120 MHz survey ($S^{\rm eff}_{1.4}/\sigma_{\rm Tier\,1}$), which 
results in an effective sensitivity of 6~mJy (see dashed lines in Fig.~\ref{fig:cluster-fractions}). 
We find that LOFAR should discover relics in more than $50$\% of the 
clusters. In the case of luminous clusters the fraction can be as high as $90$\%.

\begin{figure}
	\centering
        \includegraphics[width=0.5\textwidth]{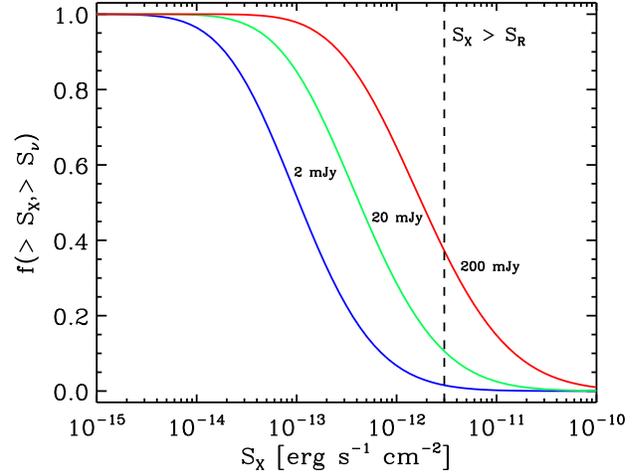}
	\caption{ 
        Same as Fig.~\ref{fig:cum_Sx} but with specifications for the LOFAR-Tier~1-120~MHz survey: 
        $\nu_{\rm obs} = 120 \: \rm MHz$ and an effective sensitivity $S^{\rm eff}_{1.4} = 20 \: \rm mJy$. 
        The curves give the cumulative number for a relic flux of 2, 20 and 200 mJy. 
        As a vertical dashed line is also shown the X-ray completeness limit of the REFLEX cluster sample, 
        $S_{\rm R}=3\times10^{-12}$ erg s$^{-1}$ cm$^{-2}$.
        }
	\label{fig:cum_Sx_LOFAR}
\end{figure}

Using Eq.~(\ref{eq:f_X}) we can also estimate what is the required sensitivity in X-ray
surveys to find at least a fraction of the relics that can be potentially discovered 
in a radio survey. Fig.~\ref{fig:cum_Sx_LOFAR} indicates 
that for relics with 20 mJy in the LOFAR-Tier 1-120 MHz 
survey, about $50$\% of the relics might be identified 
by cross-correlation if the X-ray surveys are complete up to at least 
$ 4 \times 10^{-13} \: \rm erg \, s^{-1} \, cm^{-2}$ (in doing this calculation we have 
used magnetic model `a' as done in the previous section). 
Hence, an all-sky X-ray catalogue with an X-ray flux limit one order of
magnitude below that of the REFLEX sample is necessary
to identify a considerable fraction of the relics. 

\begin{table*}
\begin{tabular}{l c r r c r r r r c c c}

\hline
\hline

Cluster             &  Type &   $S_{1.4}$   &   $\nu_{\rm obs} $    &   $z$   &    $P_{1.4}$                                          &   $\Delta_P$    &   $L_{X}$         &  $T_X$  &  NVSS  &  N$+$R  & References$^{a}$ \\
                          &            &                       &                                     &            &    $  \times 10^{24}  $                          &                             &  $ \times 10^{44} $                            &                 &          &        \\ 
                          &            &      [mJy]        &             [GHz]             &            &    [W\,Hz$^{-1}$]                                  & [$\sigma_P$]  &  [erg\,s$^{-1}$]                                  &  [keV]       &             &      \\                                                            
\hline
 0217+70                  &   C    &   -             &   -      &  0.0655    &        -    &   -    &   0.25  &        -   & \checkmark   &              & Br11\\
 0809+39                  &   C    &  62.6           &  1.4     &  -         &          -  &   -    &      -  &        -   & \checkmark   &              & BR09      \\
 1RXS 06+42               &   R    &  357.8          &  1.38    &  0.225     &     52.69   &  2.36  &   10.0  &        -   & \checkmark   &              & vWp \\
 24P73                    &   P    &  12.0           &  1.38    &  0.16      &     0.86    &   -    &      -  &        -   & \checkmark   &              & vW11a     \\
 A\,S753                  &   C    &  460.0          &  1.4     &  0.014     &     0.20    &  4.32  &   0.04  &        -   & \checkmark   &              & Su03/B04 \\
 A\,13                    &   P    &  35.5           &  1.4     &  0.0940    &     0.81    &  2.07  &   1.24  &      6.8   & \checkmark   &  \checkmark  & S01/B04 \\
 A\,85                    &   P    &  40.9           &  1.4     &  0.0555    &     0.30    &  0.39  &   5.18  &      6.4   & \checkmark   &  \checkmark  & S01/M98 \\
 A\,115                   &   R    & 14.7            &  1.4     &  0.1971    &     1.71    &  0.25  &  15.70  &      5.5   & \checkmark   &              & Go01\\
 A\,133                   &   P    & 137.0           &  1.4     &  0.0566    &     1.05    &  2.12  &   1.40  &      4.3   & \checkmark   &  \checkmark  & S01/B04 \\
 A\,521                   &   R    &    14.1         &  1.4     &  0.2475    &     2.75    &  1.10  &   7.44  &      7.0   & \checkmark   &  \checkmark  & Gi08/B04  \\
 A\,523$^{b}$             &   C    &        64.0     &  1.43    &  0.1034    &     1.70    &  2.74  &   0.89  &        -   & \checkmark   &  \checkmark  & vW11c/B00  \\
 A\,548W                  &   R    &     121.0       &  1.4     &  0.0424    &     0.51    &  3.91  &   0.11  &        -   & \checkmark   & (\checkmark) & F06/B04  \\
 A\,610                   &   C    &       18.6      &  1.4     &  0.0956    &     0.44    &  -     &      -  &        -   & \checkmark   &              & GF00\\
 A\,725                   &   C    &         6.0     &  1.4     &  0.0900    &     0.12    &  1.47  &   0.80  &        -   & \checkmark   &              & KS01/B00\\
 A\,746                   &   R    &       24.3      &  1.38    &  0.2323    &     4.05    &  1.89  &   3.68  &        -   & \checkmark   &  \checkmark  & vWp\\
 A\,754                   &   R    &         6.0     &  1.37    &  0.0542    &     0.04    & -0.39  &   3.79  &      9.0   & \checkmark   &  \checkmark  & Ma11/M11\\
 A\,781                   &   C    &       15.0      &  1.4     &  0.2952    &     4.56    &  1.83  &   4.15  &      9.9   & \checkmark   & (\checkmark) & Go11,V08,V11\\
 A\,786                   &   F    &     120.0       &  1.48    &  0.1241    &     5.31    &  2.87  &   1.53  &        -   & \checkmark   &              & GF00/B00\\
 A\,1240                  &   D    &       16.1      &  1.4     &  0.159     &     1.16    &  2.39  &   1.00  &    4.8     & \checkmark   &              & B09/D99\\
 A\,1300                  &   R    &       15.0      &  1.4     &  0.3075    &     4.89    &  0.96  &  12.12  &     13.9   & \checkmark   &  \checkmark  & R99\\
 A\,1612                  &   R    &       62.0      &  1.4     &  0.1797    &     5.80    &  2.46  &   2.41  &        -   & \checkmark   &  \checkmark  & vW11c/B04\\
 A\,1664                  &   D    &    107.0        &  1.4     &  0.1276    &     4.72    &  1.46  &   7.20  &        -   & \checkmark   &  \checkmark  & Go01\\
 A\,1758                  &   C    &      12.8       &  1.4     &  0.2799    &     3.32    &  0.86  &  10.90  &      7.0   & \checkmark   &              & G09/E98\\
 A\,2034                  &   C    &       23.6      &  1.38    &  0.1130    &     0.79    &  1.15  &   3.56  &      7.2   & \checkmark   &  \checkmark  & vW11c/E98\\
 A\,2048                  &   P    &      18.9       &  1.43    &  0.0972    &     0.47    &  1.44  &   1.90  &        -   & \checkmark   &              & vW11a/Sh08\\
 A\,2061                  &   R    &       26.7      &  1.38    &  0.0784    &     0.41    &  0.75  &   3.95  &      4.5   & \checkmark   &              & vW11/E98\\
 A\,2163                  &   R    &       18.7      &  1.4     &  0.2030    &     2.33    &  0.21  &  19.62  &     11.8   & \checkmark   &  \checkmark  & F01/B04\\
 A\,2255                  &   R    &       43.0      &  1.4     &  0.0809    &     0.70    &  1.24  &   3.08  &      6.1   & \checkmark   &  \checkmark  & PD09/E98\\
 A\,2256                  &   R    &    462.0        &  1.4     &  0.0581    &     3.96    &  1.49  &   3.69  &      6.9   & \checkmark   &  \checkmark  & CE06/E98\\
 A\,2345                  &   D    &       59.0      &  1.4     &  0.1760    &     5.35    &  2.01  &   3.91  &        -   & \checkmark   &  \checkmark  & Bo09/B04\\
 A\,2744                  &   R    &       18.2      &  1.4     &  0.3066    &     5.97    &  1.09  &  11.68  &      9.2   & \checkmark   &  \checkmark  & Go01\\
 A\,3365                  &   R    &       50.0      &  1.43    &  0.0926    &     1.12    &  2.55  &   0.86  &        -   & \checkmark   &              & vW11c/B04\\
 A\,3376                  &   D    &    302.0        &  1.4     &  0.0468    &     1.52    &  2.60  &   1.01  &      4.3   &              &  \checkmark  & Ba06/Ma98\\
 A\,3667                  &   D    &   4000.0        &  1.4     &  0.055     &     29.21   &  3.46  &   2.18  &       6.5  &              &  \checkmark  & R97/Ma98\\
 A\,4038                  &   P    &      49.0       &  1.4     &  0.0292    &     0.10    &  1.23  &   1.00  &      3.1   & \checkmark   &              & S01/B04\\
 CIZA\,0649+18            &   R    &       31.6      &  1.43    &  0.064     &     0.32    &  1.08  &   2.38  &        -   & \checkmark   &              & vW11c/E02\\
 CIZA\,2242+53            &   R    &     241.0       &  1.38    &  0.1921    &     26.05   &  2.34  &   6.80  &       5.8  & \checkmark   &              & vW10/K07,Op\\
 COMA                     &   R    &     229.2       &  1.4     &  0.0231    &     0.28    &  0.68  &   3.63  &      8.3   & \checkmark   &  \checkmark  & G91/RB02\\
 MACS\,0717+37            &   R    &     142.3       &  1.43    &  0.5548    &     200.44  &  1.98  &  32.90  &     10.5   & \checkmark   &              & vW09a,Bo09/Ed03\\
 MaxBCG\,138.9            &   C    &       24.7      &  1.4     &  0.32      &     8.87    &  -     &      -  &        -   & \checkmark   &              & vW09c\\
 MaxBCG\,J217+13          &   P    &      19.6       &  0.61    &  0.16      &     0.53    &  1.68  &   0.79  &    2.1     & \checkmark   &              & vW09c/O11\\
 PLCK\,G287.0             &   D    &       58.0      &  1.4     &  0.39      &     33.51   &  1.62  &  17.20  &   12.9     & \checkmark   &              & Ba11\\
 RXC\,1053+54             &   R    &      20.0       &  1.43    &  0.0704    &     0.25    &  2.37  &   0.44  &        -   & \checkmark   &  \checkmark  & vW11c/P04\\
 RXC\,1314-25             &   R    &      35.2       &  1.4     &  0.2439    &     6.68    &  1.31  &   9.92  &        -   & \checkmark   &  \checkmark  & V07/B04\\
 ZwCl\,0008+52            &   D    &      67.0       &  1.38    &  0.104     &     1.86    &  3.25  &   0.50  &        -   & \checkmark   &              & vW11b\\
 ZwCl\,2341+00            &   R    &      28.5       &  1.4     &  0.27      &     6.85    &  3.17  &   1.10  &      5.0   & \checkmark   &              & vW09b,G10/Ba02\\
\hline
\hline
\end{tabular}

{\small
$^{a}$Ba02:~\cite{2002NewA....7..249B}, Ba06:~\cite{bagchi:06}, Ba11:~\cite{2011arXiv1104.5551B}, Bo00:~\cite{2000ApJS..129..435B}, B04:~\cite{2004A&A...425..367B}, Bo09:~\cite{2009A&A...494..429B}, Br11:~\cite{2011ApJ...727L..25B}, BR09:~\cite{2009AJ....137.3158B}, CE06:~\cite{2006AJ....131.2900C}, D99:~\cite{1999ApJ...519..533D}, E02:~\cite{2002ApJ...580..774E}, Ed03:~\cite{2003MNRAS.339..913E}, F01:~\cite{2001A&A...373..106F}, F06:~\cite{2006MNRAS.368..544F}, Gi08: ~\cite{2008A&A...486..347G}, G91:~\cite{1991A&A...252..528G}, G09:~\cite{2009A&A...507.1257G}, G10:~\cite{2010A&A...511L...5G}, GF00:~\cite{2000NewA....5..335G}, Go01:~\cite{2001A&A...376..803G}, Go11:~\cite{2011A&A...529A..69G}, K07:~\cite{2007ApJ...662..224K}, KS01:~\cite{kempner:01}, M98:~\cite{1998ApJ...503...77M}, M03:~\cite{2003ApJ...586L..19M}, Ma11:~\cite{2011ApJ...728...82M},Op:~Ogrean et al. (in prep.), O11:~\cite{2011MNRAS.414.1175O}, P04:~\cite{2004A&A...423..449P}, PD09:~\cite{2009A&A...507..639P}, V07:~\cite{2007A&A...463..937V}, vW09a:~\cite{2009A&A...505..991V}, vW09b:~\cite{2009A&A...506.1083V}, vW09c:~\cite{2009A&A...508...75V}, vW10:~\cite{vanweeren:10}, vW11a:~\cite{2011A&A...527A.114V}, vW11b:~\cite{2011A&A...528A..38V}, vW11c:~\cite{2011A&A...533A..35V}, vWp:~van Weeren et al. (in prep.), RB02:~\cite{2002ApJ...567..716R}, R97:~\cite{roettgering:97}, R99:~\cite{1999MNRAS.302..571R}, S01:~\cite{2001AJ....122.1172S}, Sh08:~\cite{2008MNRAS.389.1074S}, Su03:~\cite{2003AJ....125.1095S}, V08:~\cite{2008A&A...484..327V}, V11:~\cite{2011MNRAS.414L..65V}.
$^{b}$\cite{2011A&A...530L...5G} has classified this source as a radio halo.
}
\vspace{-0.1cm}
\caption{List of currently known relics extracted from the literature. Columns: 
(1) Cluster name,
(2) Classification (R: single relic; D: double relic; P: phoenix; C: diffuse radio emission detected [more observations are needed to confirm its nature]; 
F: probably misclassified as relic),
(3) Radio flux (all relics in the cluster are considered),
(4) Observed frequency,
(5) Redshift, 
(6) Radio power (computed assuming a spectral index of $-1.2$),
(7) Radio power deviation (model `a'; see text),
(8) Cluster X-ray luminosity,
(9) Cluster X-ray temperature, 
(10) Checkmark if within NVSS relic sample ($\delta > -40^{\circ}$),
(11) Checkmark if within NORAS$+$REFLEX cluster sample (in brackets if below flux limit $S_{\rm R} = 3\times10^{-12}$ erg s$^{-1}$ cm$^{-2}$),
(12) References (radio/X-ray). 
}
\label{tab:relics}
\end{table*}

\begin{table*}
\begin{tabular}{ l c c c c c c c c}
\hline
\hline
Survey & $\nu_{\rm obs}$ & $\sigma_{\nu}$ & Area      & $f_{\rm s}$  & Number of relics & Number of relics & Number of relics & Number of relics \\
       & [MHz]           &       [mJy]      & [deg$^2$] &              & ($0<z<0.3$) & ($0.3<z<0.5$) & ($0.5<z<1$) & ($z>1$) \\
\hline

NVSS             & 1400           & 0.45\phantom{00} &  $\delta > -40^\circ$   & 0.82\phantom{00}  & \phantom{00}23 (26) &  \phantom{0}7 (6) &   \phantom{0}5 (3)  &  \phantom{0}0 (0) \\  
\hline 
WENSS            & \phantom{0}325 & 3.6\phantom{000} &   $\delta > 28.5^\circ$ & 0.34\phantom{00}  & \phantom{00}8 (9) &  \phantom{0}2 (2)&  \phantom{0}1 (1)    &  \phantom{0}0 (0)\\
\hline
LOFAR -- Tier 1  & \phantom{00}60 & 1.0\phantom{000}  & 20626               & 0.5\phantom{000}  & \phantom{00}260 (365) & \phantom{0}155 (170) & \phantom{0}160 (140) & \phantom{0}35 (23)  \\
                 & \phantom{0}120 & 0.1\phantom{000}  & 20626               & 0.5\phantom{000}  & \phantom{000}850 (1310) & \phantom{0}640 (800)& \phantom{0}810 (840) & \phantom{0}240 (190)  \\
                 & \phantom{0}210 & 0.065\phantom{0}  & \phantom{00}783     & 0.019\phantom{0}  & \phantom{00}27 (41) & \phantom{0}19 (24) & \phantom{0}24 (24)        & \phantom{0}7 (5)\\
LOFAR -- Tier 2  & \phantom{00}60 & 0.25\phantom{00}  & \phantom{0}1184     & 0.029\phantom{0}  & \phantom{00}46 (70) & \phantom{0}34 (42) & \phantom{0}43 (44)        & \phantom{0}12 (10)\\
                 & \phantom{0}120 & 0.025\phantom{0}  & \phantom{00}239     & 0.006\phantom{0}  & \phantom{00}27 (45)& \phantom{0}24 (34)& \phantom{0}37 (44) & \phantom{0}14 (13)\\
                 & \phantom{0}210 & 0.016\phantom{0}  & \phantom{000}78     & 0.0019            & \phantom{000}7 (12)  & \phantom{0}6 (9)& \phantom{0}10 (11) & \phantom{0}3 (3)\\
LOFAR -- Tier 3  & \phantom{0}150 & 0.0062            & \phantom{000}30     & 0.0007            & \phantom{000}7 (13) & \phantom{00}7 (11)& \phantom{0}12 (17) & \phantom{0}6 (6)\\
\hline
WODAN            & 1400           & 0.01\phantom{00}  & $\delta > 30^\circ$ & 0.33\phantom{00}  & \phantom{00}340 (510) & \phantom{0}230 (275) & \phantom{0}270 (265) & \phantom{0}70 (50)\\
                 &                & 0.005\phantom{0}  & 1000                & 0.024\phantom{0}  & \phantom{00}42 (66) & \phantom{0}32 (41)    & \phantom{0}41 (43)  & \phantom{0}12 (10)\\
\hline 
\hline
\end{tabular} 
\caption{
         Properties of various present and upcoming radio surveys. For LOFAR and WODAN different configurations are also shown. 
         The colums correspond to: name of the radio survey, observing frequency, survey noise level, observed sky area (or declination limit), corresponding sky fraction 
         and approximate number of expected relics using our set of fiducial parameters as a function of redshift for magnetic models `a' and `b' (in brackets). It is worth noting 
         that these are only plausible estimates for upcoming radio surveys but are not meant to be definitive values (see Section~\ref{sec:predictions}).
         }
\label{tab:surveys}
\end{table*}

\section{Summary}
\label{sec:summary}

  Radio relics are believed to trace merger shock fronts in galaxy
  clusters. The radio luminosity of shock fronts depends strongly on
  the Mach number of the shock, but also on the size of the front and
  on the magnetic field present in the downstream region. 
  Even if in every cluster there are shock fronts related with past 
  merger events, the actual radio luminosity caused by the shocks may vary strongly, from
  no detectable radio emission at all to the presence of luminous
  radio relics. To describe the large spread of radio luminosities
  more formally, we have introduced the {\it radio power probability
  distribution}, $p(P_\nu,M,z)$, aim at assesing the likelihood of 
  relics in galaxy clusters with a given mass and redshift in a given frequency. 
  We use the {\sc MareNostrum Universe} simulation to estimate the probability distribution. 
  To this aim, we selected the 500 most massive clusters at 5 different redshifts up to $z=1$ to
  detect shock fronts in assembling galaxy clusters. Then, we apply the
  scheme developed by \citet{2008MNRAS.391.1511H} for estimating their 
  radio relic luminosity. Based on the distribution of
  radio relic luminosities of the simulated clusters we conclude that the
  radio power probability is well approximated by a log-normal distribution. 
  Moreover, using our galaxy cluster samples, we are able to
  estimate how the radio relic distributions scale with cluster mass,
  redshift, and observing frequency.

  Using the radio power probability distribution we wish to
  determine the relic number counts, $N(>S_\nu)$. Basically this
  is given by a convolution of the probability distribution and the
  dark matter halo mass function. However, radio relics are not
  straightforwardly identified in radio observations, therefore, even
  luminous relics are possibly present in radio catalogues, but not yet
  identified as relics. For instance, the relics in 1RXS~$06$+$42$ and
  in CIZA~2242 are bright systems present in the WENSS catalogue, but have
  only recently been reported as relics. We therefore introduce the
  discovery probability as a function of radio flux, $\phi(S_\nu)$. 
  The number counts are obtained by a convolution of all
  three, the halo mass function, the radio power probability
  distribution, and the discovery probability.

  It is important to remark that it is not possible to {\it predict}
  radio flux number counts of relics purely from cosmological
  simulations. In this regard a major source of uncertainty is the
  efficiency of the electron acceleration at the shock front. We
  therefore use the observed relic number counts to determine the
  reference normalization for the radio power probability distribution.

  The resulting framework allows us to estimate the number of
  detectable relics in upcoming radio surveys. In the following we
  summarize our main conclusions:

\begin{itemize}

  \item
  To evaluate the {\sc MareNostrum Universe} simulation we {\it assumed} an
  electron acceleration efficiency, $\xi_{\rm e}=0.005$ and two different magnetic field 
  scalings with local electron density in the simulation. Normalizing the radio
  power probability distribution by the list of known NVSS relics resulted
  in lower values for the reference radio power which can be interpreted as 
  evidence for a low electron acceleration efficiency. In particular, we find that 
  $\xi_{\rm e} \lesssim 0.001$. According to the magnetic scaling proposed by \citet{2010A&A...513A..30B} 
  in the case of the Coma cluster (model `b') the acceleration efficiency could easily reach values 
  of $\xi_{\rm e} \sim 10^{-5}$. However, there are many uncertainties which may 
  affect the $P_0$ value, e.g. the actual discovery probability.
  \\

  \item
  After normalizing the radio relic number counts, $N(>S_\nu)$, we
  split the obtained number counts into redshift bins. As a result we
  expect more relics for $z>0.3$ than being observed. This might 
  indicate that the $B(n_{\rm e})$ relations assumed in the simulation 
  show an additional dependence on redshift. However, magnetic model `b' 
  seems to agree better with observational results which would point toward $\gtrsim \mu$G 
  magnetic fields in the location of radio relics. We consider this approach as a 
  very promising diagnostics of the evolution of magnetic fields in galaxy clusters 
  but larger relic samples are needed to draw any robust conclusion.
  \\

  \item
  The observed relic number counts are reasonably reproduced assuming
  an effective sensitivity of 100~mJy. Candidates for many relics may
  have been first identified in the NVSS survey, which has a r.m.s.
  noise level of 0.45~mJy. We adopt therefore that the effective
  sensitivity for finding relics is generally about 200 times the r.m.s.
  noise of a survey. We apply this to the specifications of the
  proposed LOFAR and APERTIF surveys. Under these assumptions we find that 
  the LOFAR-Tier~1-120~MHz has the potential to find more than a thousand
  radio relics.
  \\

  \item
  More than $50\%$ of the relics expected to be found with LOFAR-Tier~1-120~MHz survey
  should reside in clusters with $z>0.3$ and there should be
  even more than 100 relics in clusters with $z>1$. Hence, in principle 
  this survey will allow to discover sufficient relics to analyze the
  evolution of magnetic fields in clusters in a statistical way.
  \\

  \item
  To confidently discover a relic, the clusters which host the diffuse radio 
  emission need to be identified. Many of the relics which are 
  detectable by the LOFAR-Tier~1-120~MHz survey may reside in faint
  clusters. More precisely, we predict that about $50\%$ of the relics with
  20~mJy will reside in clusters with an X-ray flux below $ 4 \times
  10^{-13} \: \rm erg \, s^{-1} \, cm^{-2} $.
  \\

  \item
  Following \cite{2011A&A...533A..35V} we study an X-ray flux galaxy
  cluster sample based on the NORAS$+$REFLEX catalogues. About 4\% of
  the galaxy clusters in the sample host radio relics. This value is 
  significantly lower than the one obtained for the overall sample 
  since we consider only known clusters here.  We found that we can 
  reproduce the relic fraction assuming an effective sensitivity of 27~mJy at 1.4 GHz.
  As discussed in \cite{2011A&A...533A..35V} we also find that fraction of clusters 
  which host a relic increases with cluster X-ray luminosity and redshift.
  \\

  \item
  We expect that the LOFAR-Tier~1-120~MHz survey will find radio relics 
  in around 50\% of the NORAS$+$REFLEX clusters. Furthermore, for the most massive 
  clusters this fraction can be as high as 90\%.

\end{itemize}

\section*{Acknowledgments}
The authors thank Marcus Br\"uggen and Huub R\"ottgering for helpful discussions 
and careful reading of the manuscript. They also thank the anonymous referee for 
constructive comments that helped to improve this paper.
The {\sc MareNostrum Universe} simulation has been performed at the 
BSC (Barcelona, Spain) and analyzed at NIC (J\"ulich, Germany). 
SEN acknowledges support by the Deutsche Forschungsgemeinschaft 
under the grant MU1020 16-1. MH acknowledges support by the research group 
FOR 1254 ``Magnetisation of Interstellar and Intergalactic Media: The Prospects of 
Low-Frequency Radio Observations'' founded by the Deutsche Forschungsgemeinschaft.
GY acknowledges support of MICINN (Spain) through research grants FPA2009-08958, 
AYA2009-13875-C03-02 and CONSOLIDER-INGENIO SyEC (CSD2007.0050). 
He also thanks Comunidad de Madrid for partial support under 
the ASTROMADRID project (CAM S2009/ESP-1496).

\bibliography{Nuza_Hoeft_van_Weeren_et_al_2011}
\bibliographystyle{apj}

\end{document}